\documentclass[twocolumn, twocolappendix]{aastex631}

\usepackage{mathtools}
\usepackage{bm}
\usepackage{color}

\graphicspath{{./}{figures/}}

\begin{document}

\title{Spin-down of solar-mass protostars in magnetospheric accretion paradigm}

\author[0000-0003-3882-3945]{Shinsuke Takasao}
\affiliation{Department of Earth and Space Science, Graduate School of Science, Osaka University, Toyonaka, Osaka 560-0043, Japan}

\author[0000-0002-1932-3358]{Masanobu Kunitomo}
\affiliation{Department of Physics, Kurume University, 67 Asahimachi, Kurume, 830-0011, Fukuoka, Japan}
\affiliation{Universit\'e C\^ote d'Azur, Observatoire de la C\^ote d'Azur, CNRS, Laboratoire Lagrange, Bd de l'Observatoire, CS 34229, 06304 Nice cedex 4, France}

\author[0000-0001-9734-9601]{Takeru K. Suzuki}
\affiliation{School of Arts \& Sciences, The University of Tokyo, 3-8-1, Komaba, Meguro, Tokyo , 153-8902, Japan; Department of Astronomy, The University of Tokyo, 7-3-1, Hongo, Bunkyo, Tokyo, 113-0033, Japan}

\author[0000-0002-2707-7548]{Kazunari Iwasaki}
\affiliation{Center for Computational Astrophysics, National Astronomical Observatory of Japan, Mitaka, Tokyo 181-8588, Japan}

\author[0000-0001-8105-8113]{Kengo Tomida}
\affiliation{Astronomical Institute, Tohoku University, Sendai, Miyagi 980-8578, Japan}



\begin{abstract}
Stellar spin is one of the fundamental quantities that characterize a star itself and its planetary system. 
Nevertheless, stellar spin-down mechanisms in protostellar and pre-main-sequence stellar phases have been a long-standing problem in the star formation theory. 
To realize the spin-down, previous axisymmetric models based on the conventional magnetospheric paradigm have to assume massive stellar winds or produce highly time-variable magnetospheric ejections. However, this picture has been challenged by both numerical simulations and observations.
With a particular focus on the propeller regime for solar-mass stars, we propose a new picture of stellar spin-down based on our recent three-dimensional (3D) magnetohydrodynamic simulation and stellar evolution calculation.
We show that failed magnetospheric winds, unique to 3D models, significantly reduce the spin-up accretion torque, which make it easier for the star to spin-down. Additionally, the amplitude of time variability associated with magnetospheric ejections is reduced by 3D effects.
Our simulation demonstrates that the star spins down by generating a conical disk wind, driven by a rotating stellar magnetosphere. 
Our theoretical estimates, inspired by the numerical model, suggest that the conical disk wind is likely to play a crucial role in extracting stellar angular momentum during the protostellar phase.
As magnetospheric accretion is expected to occur in other accreting objects such as proto-giant planets, this study will also contribute to the understanding of the angular momentum of such objects.
\end{abstract}



\section{Introduction}\label{sec1}
Stellar spin is a key parameter to determine stellar properties of solar-mass stars, such as the level of magnetic activities.
The mechanism that regulates the stellar spin evolution has been one of the long-standing problems in the star formation theory \citep[see a review by e.g.][]{Bouvier2014}. Accreting stars receive not only mass but also angular momentum from the accretion flows, which increases the stellar angular momentum. In addition, stellar contraction via radiative cooling (the Kelvin-Helmholtz contraction) results in the stellar spin-up if the stellar angular momentum is conserved or increases in response to accretion. 
However, observations show that most of the pre-main-sequence (pre-MS) stars are rotating at speeds much lower than their breakup velocity \citep[e.g.][]{Herbst2007}, which implies that mechanisms to spin down the stars are operating effectively. The median value is approximately 10\% of the breakup velocity \citep[e.g.][]{Gallet2013}. A very similar problem is also found for planetary-mass objects \citep{Bryan2018}. Therefore, the spin-down problem is a common issue for accretion in stellar and planetary mass regimes.

Many theoretical models have been proposed to describe accretion modes that can realize the slow stellar rotation in the magnetospheric accretion paradigm. 
In the magnetospheric paradigm, the stellar field truncates the inner disk at the so-called magnetospheric radius $r_{\rm mag}$ \citep[e.g.][]{Hartmann2016}.
A widely discussed idea is the ``disk-locking" picture \citep{Ghosh_Lamb1979_paperIII,Camenzind1990RvMA,Koenigl1991}.
We define the corotation radius, $r_{\rm cor}$, as the radius at which the material in a Keplerian orbit corotates with the star.
The standard disk-locking models assume that all the stellar field lines retain a closed magnetic geometry even at larger radii of $r\gg r_{\rm cor}$. 
Stellar field lines penetrating the disk outside $r_{\rm cor}$ rotate faster than the disk, which results in the generation of the spin-down torque on the star. The star will reach a state of spin-equilibrium when the spin-up accretion torque balances with the spin-down magnetic torque \citep[see also][]{CollierCameron1993A&A,Armitage1996MNRAS}. 
In this paradigm, the spin-equilibrium is realized when $r_{\rm cor}$ is close to but larger than $r_{\rm mag}$.
To achieve the spin-equilibrium in the classical T Tauri phase, the standard models commonly require the stellar dipole field strength of $\sim 1$~kG.
Some observations found correlation between the presence of disks and the stellar spin \citep{Edwards1993AJ,Bouvier1997A&A}, which appears to be consistent with the expectation of the disk-locking scenario \citep[see also][]{Fallscheer2006,Cieza2007,Venuti2017}.

Despite some success, the standard disk-locking models have been challenged theoretically. 
\citet{Uzdensky2002ApJ} theoretically showed that a large portion of the stellar fields will inflate to open up due to the differential rotation between the star and disk, which results in the reduction of the spin-down torque \citep[see also][]{Lynden-Bell1994MNRAS}. \citet{Matt2004ApJ} examined the disk-locking theory by taking this effect into account, finding that the modified theory fails to explain the stellar spin-down.

A critical issue of the disk-locking models is the lack of considerations of mass loss processes.
Protostars generally have collimated jets with a speed similar to the escape velocity in the vicinity of the star \citep[$\sim 100~{\rm km~s^{-1}}$; e.g.][]{Ray2007}.
\citet{Shu1994} proposed an axisymmetric steady model to relate the stellar rotation and jet launching in the magnetospheric accretion paradigm, which has been called the X-wind model \citep[see also][]{Ostriker1995ApJ,Mohanty2008ApJ}. Although the X-wind model has provided great insights about the star-disk interaction, it is now recognized that the model holds many difficulties.
The X-wind model cannot be used for the general spin-down arguments because the model assumes that the star will always accrete near its disk-locked state with keeping the relation $r_{\rm cor}\approx r_{\rm mag}$. The model hypothesizes the formation of the magneto-centrifugally driven winds from the magnetospheric boundary, while no magnetohydrodynamic (MHD) simulations have found such X-winds. \citet{Romanova2009MNRAS} pointed out that the conical winds similar to X-winds can form, but the winds are non-steady and driven by a combination of the centrifugal and magnetic pressure gradient forces. \citet{Ferreira2013EAS} summarized some other critical issues in more detail.

Based on these experiences, theoretical studies have begun to carefully examine the roles of various winds and their temporal variability.
\citet{Lovelace1999} analyzed the launching of the conical winds from the magnetospheric boundary around a rapidly rotating star with $r_{\rm cor}<r_{\rm mag}$ (note the difference from the X-wind model). The regime is called a propeller regime \citep[e.g.][]{Illarionov1975,Romanova2018}. They found that the spin-down torque by the conical disk winds or jets can be comparable to the spin-up torque due to accretion, although the result depends on the property of turbulent magnetic diffusivity.
\citet{Ferreira2000MNRAS} also argued the importance of such winds in a different magnetic configuration where magnetic reconnection plays a role \citep[see also][]{Hirose1997PASJ}.
Another approach is to focus on the stellar winds.
By noting that a fraction of the stellar fields have to be open up in magnetospheric accretion \citep{Ustyugova2006}, \citet{Matt2005} proposed that the strong stellar wind powered by accretion will carry away the angular momentum along the open fields. In addition, \citet{Zanni2013} performed non-steady 2D axisymmetric MHD simulations and found that magnetospheric ejections can play a role in the stellar spin-down.

Although the above studies have greatly advanced our understanding of angular momentum flows around the star, the updated axisymmetric models still have faced some challenges. 
2D MHD simulations have found that the stellar spin-down will be possible with the help of polar jets, conical disk winds, and/or magnetospheric ejections \citep{Romanova2004ApJ,Lii2014,Romanova2018,Ireland2022ApJb}. However, the stellar accretion rate in such simulations shows significant time variability when the star is in the propeller regime, which is inconsistent with observations, 
as discussed in \citet{Gallet2019}.
The strong stellar wind may spin-down the star without producing significant time variability, but the stellar wind mass loss rate ($\dot{M}_{\rm SW}$) required for the spin-down is problematic. $\dot{M}_{\rm SW}$ of the accretion-powered stellar wind should be smaller than $\sim$1\% of the accretion rate $\dot{M}_{\rm acc}$ because of the energy constraint \citep{Cranmer2008, Zanni2011}. 
This is in contrast to the conclusion derived from 2D MHD models that the spin-down torque by the stellar wind can dominate the spin-up torque by accretion only if $\dot{M}_{\rm SW}$ is larger than approximately 10\% of $\dot{M}_{\rm acc}$ \citep{Pantolmos2020,Ireland2021}.

We consider that the challenges highlighted above stem from the inherent limitations of 2D axisymmetric models. In reality, accreting flows interact with the rotating magnetosphere through non-axisymmetric processes, such as turbulent mixing, which are not captured by axisymmetric models.
The axisymmetric models necessitate assumptions about turbulent magnetic diffusivity to simulate penetration. \citet{Ustyugova2006} demonstrated that the mass loss rate from conical disk winds varies significantly based on these model assumptions, complicating the assessment of different winds' roles in stellar spin-down.
Furthermore, axisymmetric models maintain perfect coherency in the azimuthal direction, which is an assumption not upheld in 3D cases. MHD instabilities at the magnetospheric boundary, for example, can disrupt this coherency by facilitating accretion flows that penetrate the stellar magnetosphere \citep[e.g.,][]{Kulkarni2008MNRAS}. To address these limitations, a transition to 3D modeling is necessary.

Recently, \citet{Takasao2022} (hereafter, ST22) performed 3D MHD simulations of magnetospheric accretion and presented a new picture of angular momentum transport in the magnetospheric accretion paradigm. 
We showed that the star in the propeller regime can spin down by driving a conical disk wind without showing a significant time variability in the accretion rate, which again highlights the importance of the conical disk wind.
Considering the updates provided by our 3D MHD simulations, we derive the upper limit of the spin-down time. 
This study focuses on the spin-down torque by the conical disk winds driven by the rotating magnetosphere. We take into account the stellar evolution and estimate the upper limit of the spin-down time at each age.
We show that the 3D effects of star-disk interaction are key to resolving the challenges of the stellar spin-down.

\section{General picture of accretion and ejection}
\subsection{A brief review of 2D models}

\begin{figure*}
\centering
\includegraphics[width=2\columnwidth]{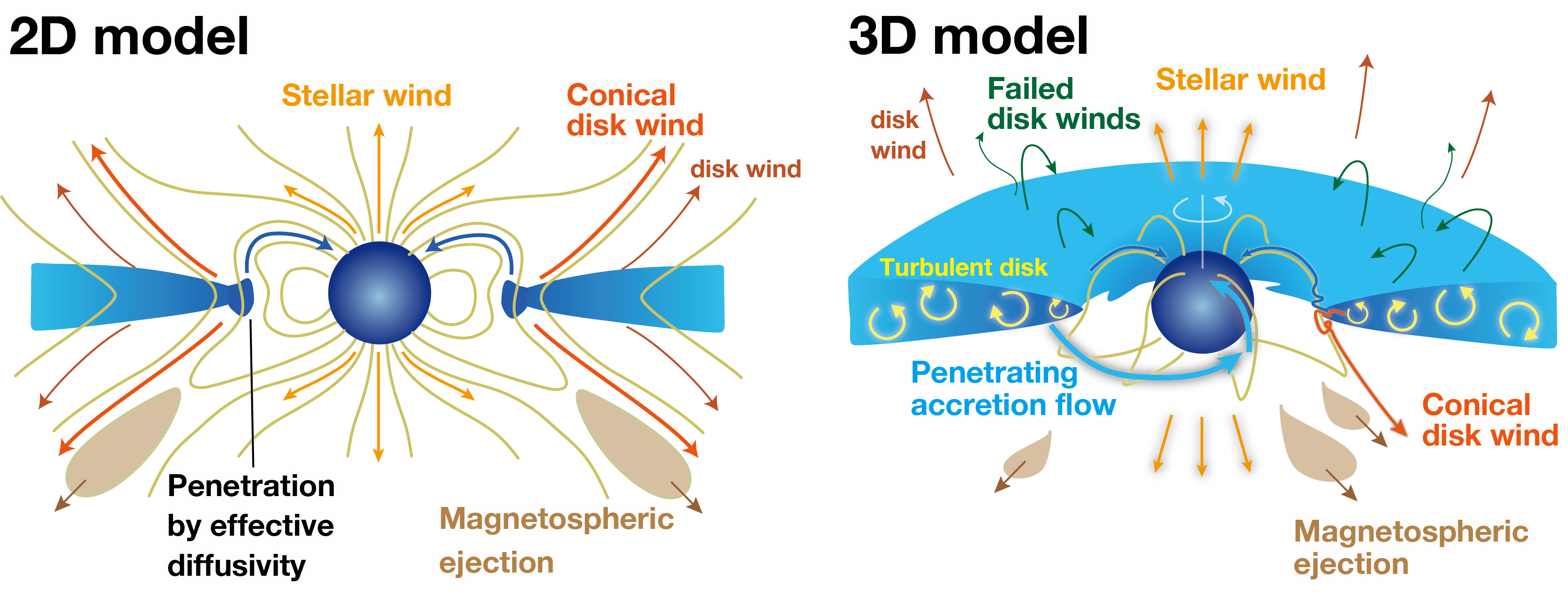}
\caption{Schematic illustration of the accretion and ejection structures in 2D (left) and 3D (right) models. The illustration for the 3D model is based on the results of ST22.}\label{fig:summary_acc_ejec}
\end{figure*}

As we are interested in the stellar spin-down, we only consider the propeller regime. We first review general properties of 2D models and then overview key results of the 3D MHD simulation by ST22, with a particular focus on the 3D effects.

Previous 2D studies predict the following types of ejections \citep[e.g.][]{Romanova2018}:
\begin{enumerate}
    \item Stellar winds
    \item Magnetospheric ejections
    \item Disk winds
    \begin{enumerate}
        \item Conical disk winds driven by the star-disk interaction
        \item Disk winds driven by the disk fields
    \end{enumerate}
\end{enumerate}
The left panel of Figure~\ref{fig:summary_acc_ejec} describes the above structures. The stellar winds blow from the polar regions. If the protostar is in the strong propeller regime ($r_{\rm cor}\ll r_{\rm mag}$), the magnetically driven polar jet will also appear \citep{Romanova2005}. Magnetospheric ejections are plasma ejections associated with magnetic reconnection of the stellar magnetosphere \citep[e.g.][]{Hayashi1996ApJ,Zanni2013}. They occur when the stellar magnetic fields are sufficiently twisted by the differential rotation between the protostar and disk gas \citep[e.g.][]{Lynden-Bell1994MNRAS}. Note that in the 2D models, the reconnection region extends in the azimuthal direction as a ring owing to the axisymmetry. As a result, the magnetospheric ejections take a form of a torus-like structure. In the magnetosphere-disk interface, disk gas penetrates in response to an effective diffusivity that imitates the turbulent diffusion. A fraction of the penetrating flow accretes onto the protostar, while the rest of it is expelled away as the conical disk wind by the rotating stellar magnetic fields. The conical disk winds often shows a north-south asymmetric structure \citep[e.g.][]{Lii2014}, where accretion occurs in a hemisphere and ejections can proceed in the other hemisphere.
The conical disk wind will be surrounded by the disk wind driven by the disk fields.

In the propeller regime, the protostar is expected to expel the angular momentum to the disk gas via the magnetosphere-disk interaction. However, a quantitative conclusion is difficult to draw due to some assumptions.
Previous studies investigated the mass and angular momentum transfer by commonly assuming that the magnetosphere is nearly rigidly rotating with the protostar \citep[e.g.][]{Lovelace1999,Ustyugova2006,DAngelo2010}. However, the rotation profile of the magnetosphere depends on the details of the twisting of the magnetosphere through the magnetosphere-disk interaction \citep[e.g.][]{Kluzniak2007ApJ}.
We also note that the bifurcation of the accretion and ejection is sensitive to the adopted diffusivity model \citep{Ustyugova2006}. 
The mass loading to the magnetosphere in the 2D models occurs through the effective diffusivity which imitates the turbulent mixing \citep[e.g.][]{Shu1994} (we note that magnetic reconnection between the magnetosphere and disk fields can provide another path of mass loading \citep{Hirose1997PASJ,Ferreira2000MNRAS}).
However, the details about (effective) magnetic diffusivity remain poorly understood.
Therefore, it is necessary to examine the assumption about the rigid rotation and the property of the turbulent mixing using 3D MHD simulations.

The accretion torque $\dot{J}_{\rm acc}$ is commonly approximated as \citep[e.g.][]{Matt2005}
\begin{align}
    \dot{J}_{\rm acc}'=\dot{M}\sqrt{GM_\ast r_{\rm mag}}
\end{align}
where $\dot{M}$ is the rate of mass accretion onto the protostar, $G$ is the gravitational constant, and $M_\ast$ is the protostellar mass. The above estimate is based on the assumption that the accreting flows bring the angular momentum which they have at the magnetospheric boundary to the protostar.
However, 2D MHD simulations find that the winds emanating from the magnetospheric boundary carry away a part of the angular momentum of the accretion flows \citep{Zanni2013,Ireland2021}. Therefore, it is reasonable to use the expression of $\dot{J}_{\rm acc}=K_{\rm acc}\dot{J}_{\rm acc}'$ for a more realistic accretion torque, where $0 \le K_{\rm acc} \le 1$.
Previous 2D models find a reduction of the spin-up torque by a few 10\%: $K_{\rm acc}\approx 0.7-0.8$ \citep{Zanni2013,Ireland2021}.

2D models predict that accretion in protostars in the propeller regime may cease and occur in a episodic way, which seems to be inconsistent with observations.
Although the star-disk interaction in 2D depends on the adopted diffusivity models, \citet{Ustyugova2006} suggest that powerful outflows can almost quench accretion in the propeller regime. However, protostars systematically show higher accretion rates than T Tauri stars \citep[e.g.][]{Fiorellino2021A&A}, which raises a question about the accretion quenching.
Time-variable accretion is also a common result of 2D models. The accretion rate can change by an order of magnitude as a result of magnetospheric ejections \citep{Romanova2004ApJ,Lii2014}. \citet{Zanni2013} and \citet{Ireland2022ApJb} clarify the importance of magnetospheric ejections for the stellar spin-down, and some observations indeed find indications of ejections \citep{Bouvier2023A&A}. However, observations do not commonly find such strong variability, challenging our understanding based on 2D models. The above challenges motivate us to perform 3D simulations.

\subsection{A overview of 3D model of ST22}\label{subsec:overview_ST22}

\begin{figure}
\centering
\includegraphics[width=\columnwidth]{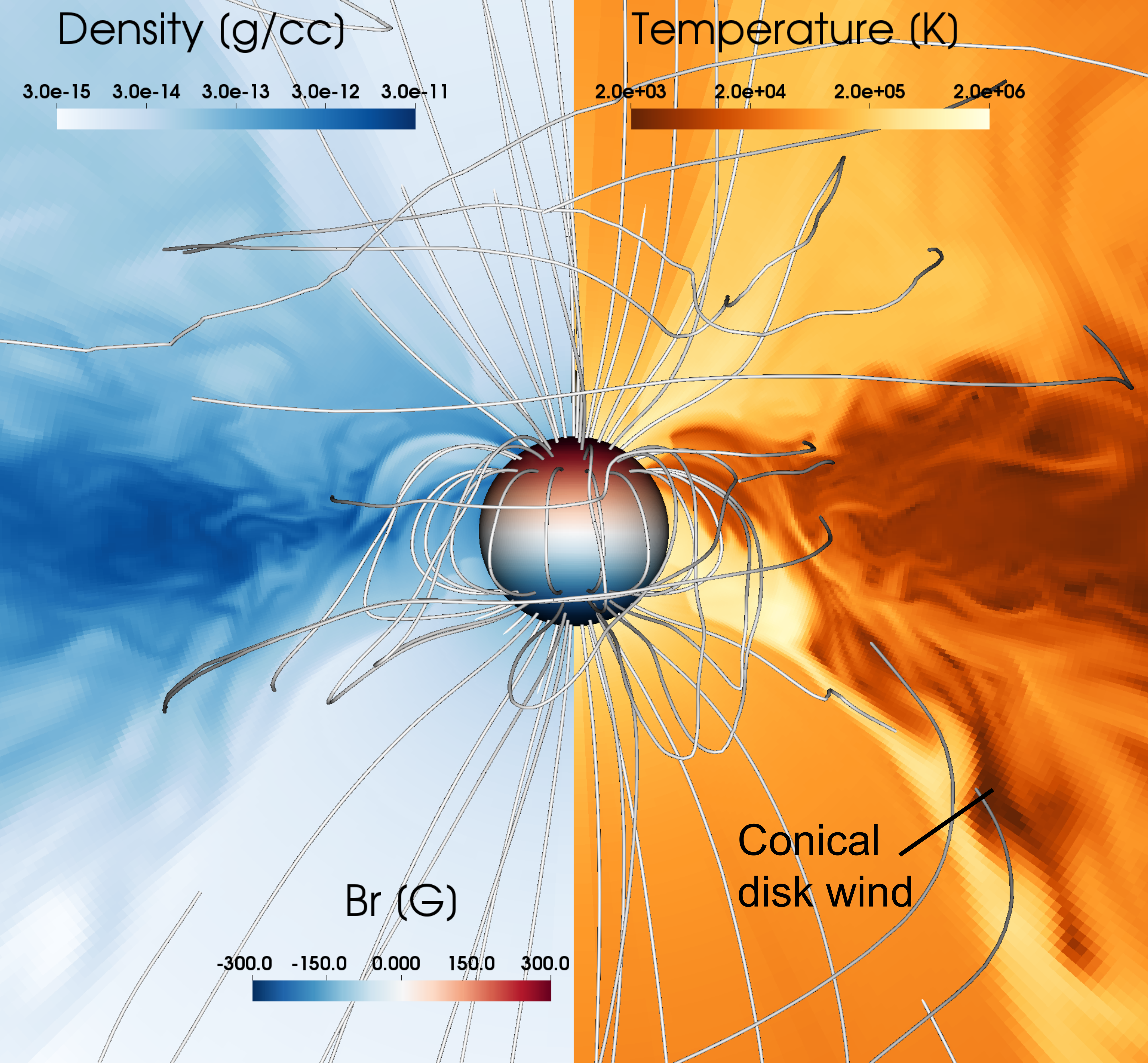}
\caption{Accretion and wind structures of the 3D MHD simulation (Model A of ST22. $r_{\rm cor}=1.5R_*$ and $r_{\rm mag}\approx 2.5R_*$). The left and right panels show the density and temperature cutouts, respectively. Lines denote magnetic field lines. The stellar surface is colored with the radial component of magnetic fields.}\label{fig:3D}
\end{figure}

\begin{figure*}
\centering
\includegraphics[width=2\columnwidth]{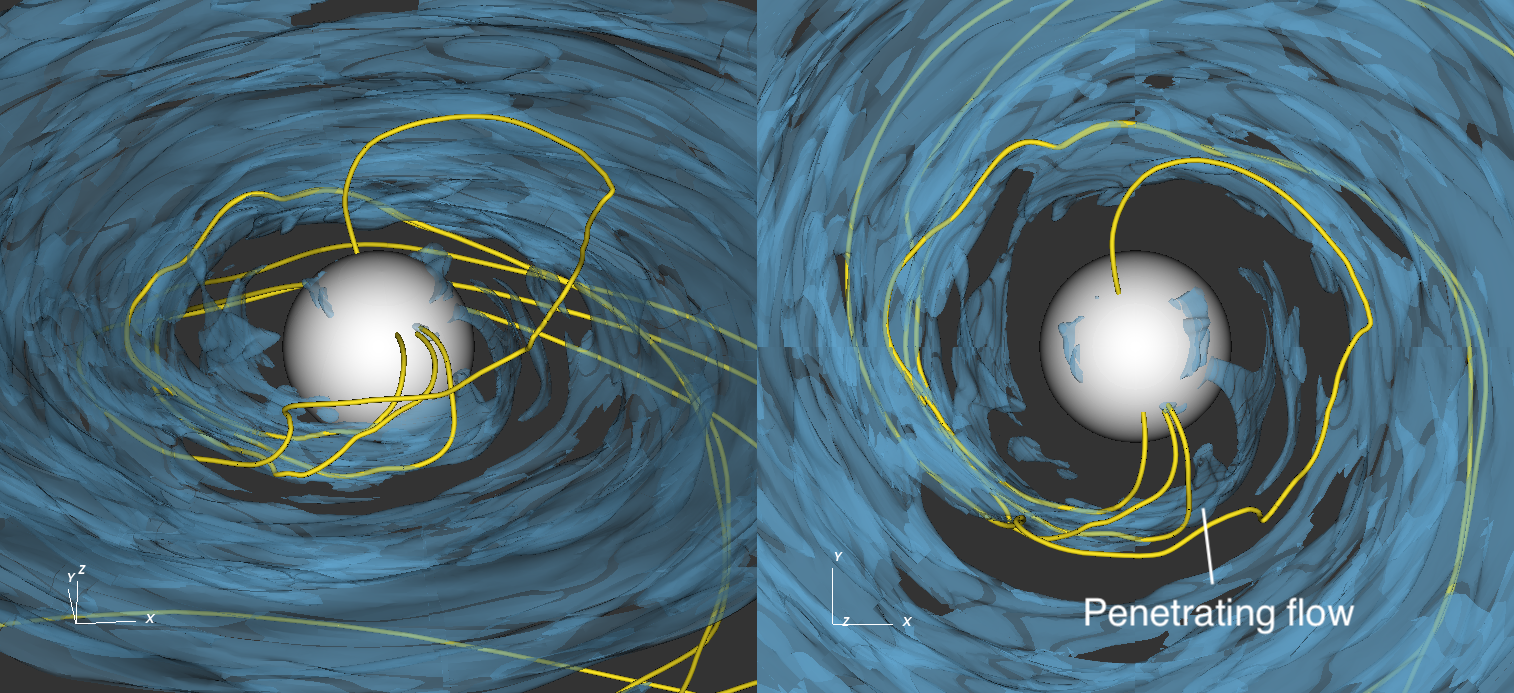}
\caption{An example of filamentary accretion flows penetrating into the magnetosphere ($t=151.1$~day of Model~A of ST22). The structure is viewed from two different points of view. The density isosurface is colored in blue. Four field lines threading the penetrating flow is denoted as yellow lines.}\label{fig:penetrating_flow}
\end{figure*}

Figure~\ref{fig:3D} presents the snapshot of our 3D simulation (Model~A of ST22). The accretion disk is turbulent in response to magneto-rotational instability \citep[MRI][]{Balbus1991ApJ}. Therefore, we can study the magnetosphere-disk interface without assuming an effective diffusivity, although we still require convergence check of the results by performing higher spatial resolution simulations.
The ejections found in previous 2D models also appear in the model: the stellar winds, magnetospheric ejections, and asymmetric conical disk winds. The 3D model further exhibits additional types of winds. A key distinction between the 2D and 3D models lies in the presence of turbulence and the non-uniform structures in the azimuthal direction. 
Asymmetric jets are indeed commonly observed in young protostars \citep{Podio2021A&A}, which seems to provide observational supports to our model.

In three-dimension, the mass loading to the magnetosphere is mediated not only by the turbulent mixing but also the penetration of the filamentary flows. Figure~\ref{fig:penetrating_flow} presents an example of filamentary accretion flows penetrating into the magnetosphere. A schematic illustration is given in the right panel of Figure~\ref{fig:summary_acc_ejec}. 
It has been recognized that similar unstable magnetospheric accretion occurs in slowly rotating stars as a result of magnetic Rayleigh-Taylor instability \citep{Kulkarni2008MNRAS,Blinova2016MNRAS}, but the stability in the propeller regime was unclear. ST22 showed that penetrating accretion flows can also form even in the case of the propeller regime but in response to different instabilities (probably the magneto-gradient driven instability; \cite{Hirabayashi2016}).
The difference in the stellar spin rate appears in the thickness of filaments: filaments in the propeller regime are thicker than those in slow rotators. We can understand the result as that the velocity shear in the magnetosphere smears out their small-scale structure.
The formation of penetrating flows prevents the accumulation of mass at the magnetospheric edge and suppresses the amplitude of time variability in accretion, which is distinct from the 2D models.

The penetrating flows are dragging the disk toroidal fields, which allows them to continuously transport their angular momentum to the disk gas. Figure~\ref{fig:penetrating_flow} displays some field lines threading the flow. The field lines go through the midplane and extend to the other side of the disk. 
Magnetic fields in the penetrating flows are connected with the protostar probably because they experience magnetic reconnection with the stellar field.
Note that the field lines are trailing with respect to the stellar rotation, which indicates that the penetrating flow is efficiently losing the angular momentum. The trailing fields also exert the spin-down torque onto the protostar.

The penetrating flows are often accompanied by slowly outgoing flows as a back-reaction of accretion. The radial speed of the accelerated gas is considerably smaller than the escape velocity. Therefore, the outflows will fail to escape from the stellar gravity. We categorize them as a type of ``failed magnetospheric winds".
In ST22, we describe them as a type of turbulent winds because of their disturbed structures. However, as the penetrating flows have coherent structures, they are not necessarily turbulent.
A more detailed analysis of the angular momentum flows is presented in Appendix~\ref{app:angular_mom_flows}.

We identify another type of failed magnetospheric winds, which is associated with the MRI turbulence.
3D MHD simulations have commonly found turbulent disk winds above MRI-turbulent disks \citep[e.g.][]{Suzuki2009ApJ,Bai2013ApJ,Takasao2018}. As their acceleration is inefficient due to their turbulent nature, most of the winds fail to escape (they are indicated as green arrows in Figure~\ref{fig:summary_acc_ejec}). When they fall, they continuously shear the magnetic field and increase the Maxwell stress. The continuous growth of the Maxwell stress leads to a runaway removal of angular momentum from the accreting material.
As a result, failed winds can form patchy accretion streams with a velocity similar to the free-fall velocity \citep{Takasao2018}. The mechanism of the angular momentum loss is similar to the onset of MRI and the other type of the failed magnetospheric winds. See also \citet{Zhu2018ApJ,Jacquemin-Ide2021A&A} for similar flows at larger scales.
The failed disk winds drive outflows when they fall as a back-reaction of angular momentum loss.
ST22 demonstrated that the failed winds also form just around the turbulent magnetosphere-disk interface. The failed winds hit the magnetosphere and become a part of the magnetospheric funnel accretion flows.

\begin{figure}
\centering
\includegraphics[width=\columnwidth]{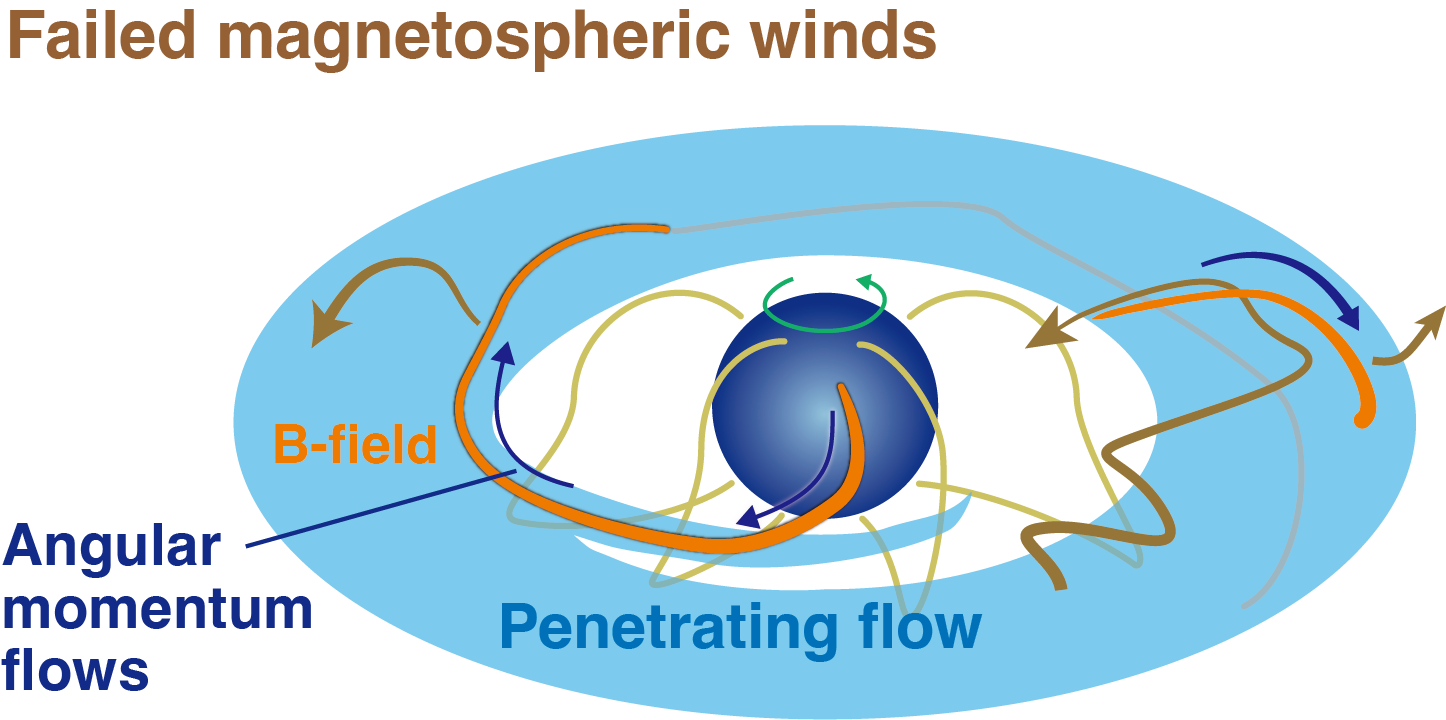}
\caption{A schematic illustration of two types of failed magnetospheric winds. Orange lines denote magnetic field lines associated with failed winds. Brown lines with arrows indicate failed winds. Dark blue arrows show the flow of angular momentum along field lines.}\label{fig:summary_failed_winds}
\end{figure}

Figure~\ref{fig:summary_failed_winds} presents a schematic illustration of failed magnetospheric winds. The two types of winds are shown: the failed winds associated with a penetrating flow and the failed winds emanating from the turbulent magnetosphere-disk interface. Also see Figure~\ref{fig:penetrating_flow} regarding the field structure of penetrating flows. The difference between the two types of winds lies in the mechanisms to produce the accretion streams. The penetrating flows are created by MHD instabilities at the magnetospheric boundary, while the patchy structure of the winds from the magnetospheric boundary is formed by the disk turbulence.
In both cases, accreting flows experience the runaway angular momentum loss by twisting the magnetic fields. As a result, the rate at which angular momentum is injected into the protostar is significantly lower than the classical estimation ($\dot{J}_{\rm acc}^\prime$). The simulation suggests that $K_{\rm acc}=0.1$ is a reasonable choice. 
The value may evolve in response to the stellar and disk evolution, which needs to be investigated in future studies.

The rotation profile of the magnetosphere is found to be different from the expectations based on 2D models because of the presence of penetrating flows (Section~3.4 of ST22). The penetrating flows which form around the midplane have a substantial inertia. After penetration, the dense flows persist in rotating for a few orbits before ultimately descending to the protostar, despite the effects of magnetic braking. As a result, they force to rotate a large body of the magnetosphere nearly at the Keplerian velocity. Therefore, the assumption of the rigid rotation inside the magnetosphere is found to be invalid. Rather, the Keplerian rotation in the magnetosphere-disk interface seems to be a better assumption (the assumption of the rigid rotation will be valid in the case that the accretion rate is so small that penetrating flows cannot affect the rotational profile of the magnetosphere). 
\citet{Kluzniak2007ApJ} also investigated a smooth transition of the rotational profile from the Keplerian disk to the rotating star. Their analytical solutions indeed show a similar near-Keplerian rotation in the outer magnetosphere. 
However, their analytical solutions do not match the result of our 3D models probably because their analytical model lacks 3D effects such as the forced rotation by penetrating flows and the vertical transport of angular momentum by the stellar fields.

The role of magnetospheric ejections in the stellar spin-down seems to be limited in 3D, although they play critical roles in 2D models.
Our 3D model does not show significant time variability in the accretion rate even though magnetospheric ejections occur, which indicates that magnetospheric ejections are not as violent as found in 2D models. The difference is related to the penetrating flows. In 2D models, the stellar magnetosphere can be coherently twisted by the rotating gas that accumulate at the magnetospheric boundary. However, in 3D the gas does not accumulate at the magnetospheric boundary but forms penetrating flows. The penetrating flows cannot twist the magnetosphere coherently because they only twist a part of the stellar magnetosphere. The resulting magnetospheric ejections are patchy in the azimuthal direction (Figure~\ref{fig:summary_acc_ejec}), which results in a weaker energy release than in 2D cases. See ST22 for more reasons why powerful magnetospheric ejections are difficult to occur in 3D.

We also note that, in Model A of ST22, distinguishing between magnetospheric ejections and conical disk winds is difficult in many cases.
The conical disk wind itself shows a significant density inhomogeneity (Figure~\ref{fig:3D}) probably because the turbulent disk gas is accelerated by fluctuating fields. In addition, magnetospheric ejections with coherent structures are uncommon.
For these reasons, we include magnetospheric ejections as a part of the conical disk winds in our analysis. We consider that this treatment will not significantly affect our arguments about the conical disk winds as long as the time variability is insignificant.

In summary, the nonuniform magnetospheric structure in the azimuthal direction (including turbulence) is key to explain the spin-down of protostars showing a low level variability. The nonuniform structure allows accretion and ejection to occur simultaneously. Because of the ``traffic control" of mass and angular momentum flows, the 3D model shows a much lower level of variability than 2D models. Accretion flows around the magnetosphere take filamentary or patchy structures. They lose a large fraction of angular momentum during the fall in a runaway manner, which results in the reduction of the spin-up torque by accretion. The filamentary penetrating flows force to rotate the magnetosphere nearly at the Keplerian speed, which indicates that the Keplerian rotation of the magnetosphere is a more realistic assumption than the rigid rotation around the magnetosphere-disk interface. 
Torus-like magnetospheric ejections found in 2D models are rare in 3D because of the localization of magnetic reconnection regions in the azimuthal direction, which suggests that their role in stellar spin-down should be limited in 3D. Therefore, we focus on the spin-down torque by conical disk winds.

\section{Modelling spin-down torque by conical disk winds}\label{sec:cdw}
In general, the spin-down torque by a magnetically driven wind can be estimated as follows:
\begin{align}
    \dot{J}_{\rm wind} \approx \dot{M}_{\rm wind}r_A^2 \Omega(r_0), \label{eq:general_Jdot}
\end{align}
where $\dot{M}_{\rm wind}$ is the wind mass loss rate, $r_{\rm A}$ denotes the Alfv\'en radius, and $\Omega(r_0)$ is the angular velocity of the magnetic field at the radius of the wind base $r=r_0$ \citep[e.g.][]{Pascucci2023ASPC}. 
The Alfv\'en radius indicates the size of the so-called Alfv\'en surface, where the poloidal velocity of the plasma flow is equal to the poloidal Alfv\'en velocity.
The magnetic field inside the Alfv\'en surface nearly rigidly rotates with the star.
To calculate the spin-down torque, we need to estimate each component on the right-hand side of Equation~(\ref{eq:general_Jdot}).

We consider a more specific functional form of Equation~\ref{eq:general_Jdot} for the conical disk winds.
Considering that a large portion of the magnetosphere rotates nearly at the Keplerian velocity because of the penetrating accretion flows, the rotation rate of the magnetosphere at the magnetosphere-disk interface will be nearly the Keplerian value there, $\Omega_{\rm K}(r_{\rm mag})$. Namely, $\Omega(r_0)=\Omega_{\rm K}(r_{\rm mag})$, which is distinct from the common assumption of the rigid rotation, $\Omega(r_0)=\Omega_\ast$.
This eliminates the explicit dependence of the stellar spin rate from the spin-down torque (Equation~(\ref{eq:general_Jdot})).

The Alfv\'en radius $r_A$ for the conical disk winds will be comparable to or larger than the magnetospheric radius because the mass loading mainly occurs at $r=r_{\rm mag}$. Indeed, it is a few of $r_{\rm mag}$ in Model~A of ST22 (see Appendix~\ref{app:Alfven_surface}). To clarify the relation between the two, we write $r_A$ as $r_A=f_{A}r_{\rm mag}$, where $f_{A}$ is a nondimensional value that is larger than unity. The functional form of $f_{A}$ will depend on relevant quantities such as the stellar spin rate and mass loss rate. In this study, we follow the suggestion by \citet{Ferreira2000MNRAS} to consider the possible range of $f_A$.
Considering observations of jets and analytical solution of disk winds based on \citet{Ferreira1997A&A}, \citet{Ferreira2000MNRAS} inferred that the magnetic lever arm $\lambda$, which is defined as $r_A^2 \approx \lambda r_{\rm cor}^2$, will be in the range of $2\lesssim \lambda \lesssim 7$. We note the relation $f_A^2/\lambda =(r_{\rm cor}/r_{\rm mag})^2$.
The range corresponds to $1\lesssim f_A \lesssim 3$ in the case of a weak propeller regime ($r_{\rm mag}\approx r_{\rm cor}$). The value of $f_A$ may be close to unity in the case of a strong propeller regime owing to a high mass loss rate. Despite the uncertainties about the magnetic lever arm, the value of $f_A$ seems to be limited in the small range. Therefore, in this study we assume that the value can be approximated as a constant when the protostar drives the conical disk winds and we adopt $f_A=2$ as a fiducial value. This assumption needs to be examined in future studies.

The spin-down torque by the conical disk wind driven from the magnetospheric boundary can then be estimated as
\begin{align}
    \dot{J}_{\rm CDW} \approx f_A^2 \dot{M}_{\rm CDW} r_{\rm mag}^2 \Omega_{\rm K}(r_{\rm mag}), \label{eq:torque_CDW}
\end{align}
where $\dot{M}_{\rm CDW}$ denotes the mass loss rate of the conical disk wind. We introduce the wind mass loss efficiency $f_{\rm eff}$ such that $\dot{M}_{\rm CDW}=f_{\rm eff}\dot{M}_{\rm acc}$, where $0 < f_{\rm eff} < 1$ and $\dot{M}_{\rm acc}$ is the accretion rate onto the star. In Model A of ST22, $f_{\rm eff}\approx 0.1-0.2$, which seems to be similar to the observational estimations of the efficiency \citep[e.g.][]{Ray2007}.
The efficiency will be determined by the details of the mass loading to the magnetosphere, which will depend on the property of turbulence and the stellar and disk parameters. Nevertheless, we theoretically infer that  $\dot{M}_{\rm CDW}\propto \dot{M}_{\rm acc}$ and $f_{\rm eff}=\mathcal{O}(0.1)$, as argued in Appendix~\ref{app:mass_loss_rate}. Considering the numerical result, we adopt $f_{\rm eff}=0.2$ as a fiducial value here. 
We also compare the mass loading between the stellar wind and the conical disk wind in Appendix~\ref{app:comparison_mass_loading} to emphasize the importance of the conical disk winds.

This study uses Equation~(\ref{eq:torque_CDW}) to estimate the upper limit of the spin-down time of the protostar in the propeller regime, by considering that the conical disk winds are the main carrier of the stellar angular momentum. ST22 showed that the conical disk winds are continuously blowing in the propeller regime, while they are intermittent and sometimes absent in the non-propeller regime (Model~B and C in ST22), which implies that $f_{\rm eff}$ depends on the stellar spin. This result limits our discussion in the propeller regime only. This study further assumes that the protostar will reach a spin-equilibrium once it decelerates to a point where $r_{\rm cor}\approx r_{\rm mag}$ as a result of the balance between the spin-up and spin-down torques. \citet{Long2005ApJ} suggested that in 2D simulations, $r_{\rm cor}\approx f_{\rm cor,eq}r_{\rm mag}$ in the spin equilibrium state, where $f_{\rm cor,eq}=1.3\text{-}1.5$ \citep[see also][]{Bouvier2007prpl}. We note that the value of $f_{\rm cor,eq}$ might differ in 3D models. The details of achieving a spin-equilibrium, however, is beyond the scope of this study.
Considering the uncertainties in the current understanding, we choose $f_{\rm cor,eq}=1$ for calculating the angular momentum of the protostar in a spin-equilibrium.

The protostar in the spin-equilibrium may re-enter the propeller regime by spinning-up in response to the disk and stellar evolution. In this case, we consider that the protostar will again start to blow the continuous conical disk winds and spin down itself. We also estimate the spin-down timescale for this case by assuming that the protostellar spin rate is close to the value in the spin-equilibrium.

\section{Stellar evolution model}
To compute the spin-down timescale across different stellar ages, it is necessary to calculate key stellar parameters, such as the radius and moment of inertia, for an accreting star. We utilize the MESA code to simulate the evolution of a young accreting star to facilitate these calculations. However, it is important to note that this study is designed to estimate the upper limit of the spin-down time at various stellar ages. For simplicity, we employ the MESA solutions that do not account for the effects of stellar rotation as references, thereby omitting the calculation of long-term spin evolution. The assumptions regarding the stellar magnetic field, such as the stellar dynamo and the removal of fossil fields, will influence the initial conditions for spin evolution \citep[e.g.,][]{Takasao2019}. Additionally, the size of the stellar magnetosphere, which may vary with the disk’s magnetic field, remains poorly understood \citep[e.g.,][]{Ferreira2000MNRAS}. Given these complexities, a detailed analysis of long-term spin evolution is reserved for future studies.
This section provides brief explanations of our methods. For more detailed descriptions, readers are referred to Appendix\,\ref{app:starevol}.

The accretion rate $\dot{M}_{\rm acc}$ is a function of time (stellar age) $t_{\rm age}$. We assume a constant rate of $\dot{M}_{\rm acc}=10^{-5}~M_\odot~{\rm yr^{-1}}$ until $3.1\times10^4\,$yr.
After that, it decreases as 
\begin{align}
    \dot{M}_{\rm acc}\propto t_{\rm age}^{-a},\label{eq:mdot_acc}
\end{align}
where $a>0$ is a measure of the rapidness of the decrease in the accretion rate. Considering the estimation by \citet{Hartmann1998}, we adopt $a=3/2$ as a fiducial value. 
The resulting final mass of the star is $1\,M_\odot$, which is achieved at $t=10$~Myr. We have confirmed that our conclusions are insensitive to the choice of the index $a$ (Appendix\,\ref{app:starevol}).

We assume that the dipole magnetosphere is established in a sufficiently early phase (possibly, $\lesssim 0.1$~Myr).
The magnetospheric radius $r_{\rm mag}$ is calculated using the equation derived in ST22, a modified version of the Ghosh \& Lamb relation \citep{Ghosh_Lamb1979_paperIII}:
\begin{align}
    \frac{r_{\rm mag}}{R_\ast} &= \left(\frac{\eta'^2 \mu_*^4}{4GM_*\dot{M}^2}\right)^{1/7} \nonumber \\
    &\approx 6.6 \left(\frac{\eta'}{1}\right)^{2/7}\left( \frac{B_*}{1~{\rm kG}}\right)^{4/7}\left(\frac{R_*}{2R_\odot}\right)^{5/7} \nonumber \\
    &\times   \left(\frac{M_*}{0.5M_\odot}\right)^{-1/7}\left(\frac{\dot{M}}{10^{-8}M_\odot~{\rm yr^{-1}}}\right)^{-2/7},\label{eq:rmag}
\end{align}
where $\eta'$ denotes a twist of magnetic fields. The detailed dependence of $\eta'$ on the thermal property of the disk gas needs to be investigated in future studies, but we set the value to unity for a conservative discussion.
ST22 derived the above equation very similar to the Ghosh \& Lamb relation using the angular momentum transfer equation.
We note that the magnetospheric radius in this study is defined as an azimuthally averaged quantity for the disturbed magnetosphere \citep[see also][]{Blinova2016MNRAS, Takasao2022}. Equation~\ref{eq:rmag} approximately agrees with the results of ST22.
Observations indicate that CTTSs generally host kilo-Gauss surface magnetic fields \citep[e.g.][]{Johnstone2014}. The field strength is comparable to or larger than the field strength determined by the pressure balance between the surface gas pressure and the magnetic pressure \citep{Safier1999,Johns-Krull2007}. Considering the observations, we adopt 1-2~kG as a typical value of the stellar field. 
We assume that the above formula is valid in the propeller regime.

The dipole field assumption seems reasonable, particularly for young and rapidly rotating T Tauri stars \citep{Gregory2012}. 
Magnetic obliquity will affect the inner disk structure when the inclination angle is greater than a few $10^\circ$ \citep{Romanova2003ApJ}. However, observations suggest that the majority of the stars show an inclination angle of $\sim 10^\circ$ only \citep{McGinnis2020}. Therefore, we ignore the effects of misalignment and discuss the spin-down process based on our 3D model ST22 where the axis of the dipole field is aligned with the rotation axis.

When we calculate the stellar angular momentum, we take two angular speeds as references. One is the maximum stellar spin, and the other is the value for a spin-equilibrium state. \citet{Lin_MinKai2011} suggest that gravitational torques prevent the protostar from spinning up to more than half of its breakup velocity.
Considering their result, we take the maximum angular velocity of the protostar, $\Omega_{\rm \ast, max}$, as $0.5\Omega_{\rm br}$, where $\Omega_{\rm br}=\sqrt{G M_*/R_*^3}$ is the break-up angular speed.
When the protostar is in the spin-equilibrium ($r_{\rm cor}=f_{\rm cor,eq}r_{\rm mag}$ and $f_{\rm cor,eq}\approx 1$), the stellar angular velocity is estimated to be $\Omega_{\rm \ast, eq}=f_{\rm cor,eq}^{-3/2}\Omega_{\rm K}(r_{\rm mag})$. In this study, to define the propeller regime, we simply assume that $f_{\rm cor,eq}=1$ in a spin equilibrium. The detailed condition about spin equilibrium is beyond the scope of this study.

\section{Upper limit \\
of the spin-down time}

\begin{figure*}
\centering
\includegraphics[width=2\columnwidth]{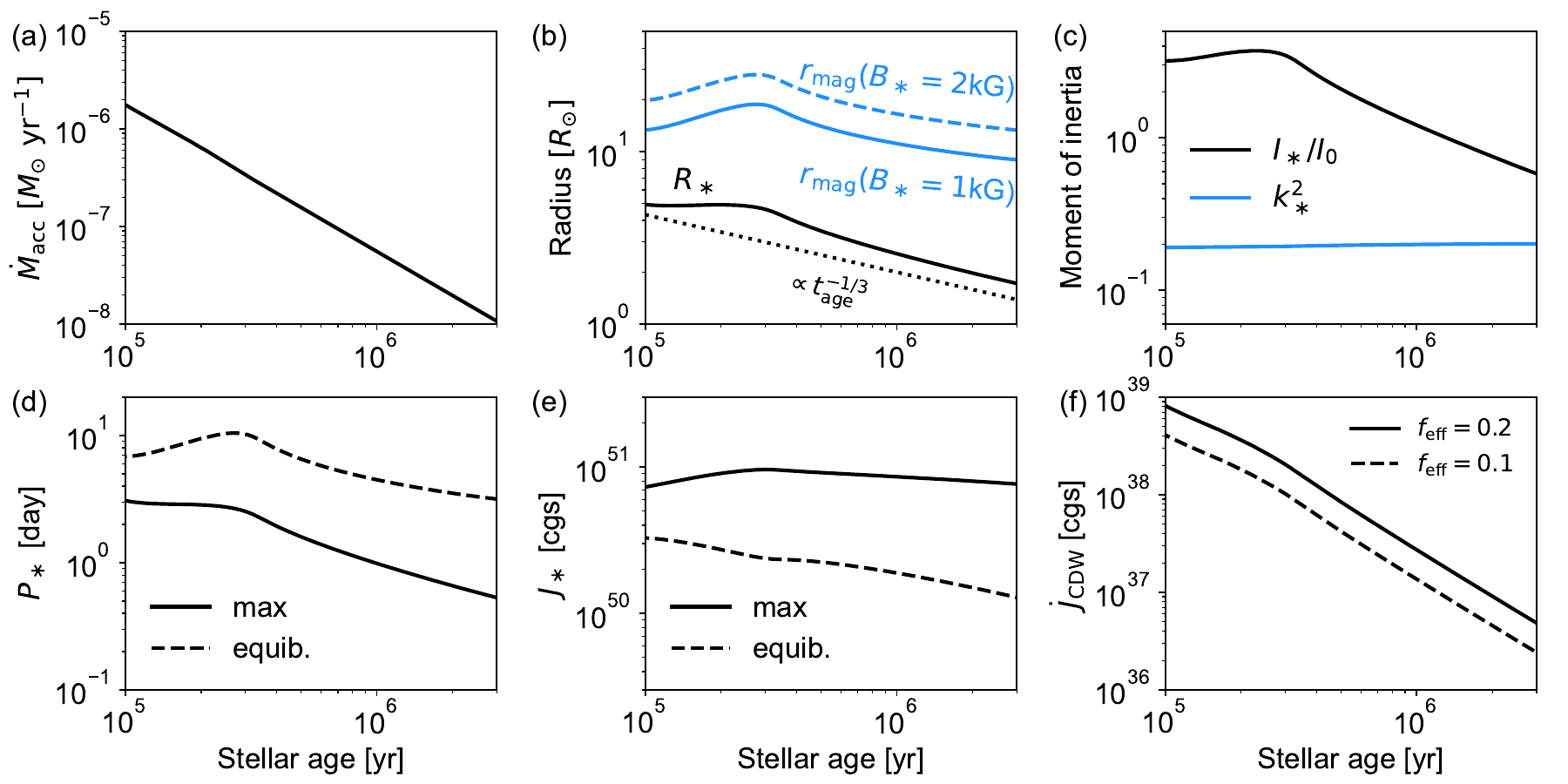}
\caption{Evolution of key quantities. Panel (a): the stellar accretion rate. Panel (b): the stellar radius (black solid line) and the magnetospheric radius (blue solid and dashed lines). Panel (c): the moment of inertia. The black and blue lines show $I_*/I_0$ and $k_*$, respectively. Panel (d): the rotational periods for a maximum rotation (solid) and in a hypothetical spin-equilibrium state (dashed), where $r_{\rm cor}=r_{\rm mag}$ is assumed. Panel (e): the stellar angular momenta for a maximum rotation (solid) and in the hypothetical spin-equilibrium state (dashed). Panel (f): the spin-down torque by the conical disk wind ($B_\ast =1$~kG).}\label{fig:general_evolution}
\end{figure*}

We first describe the general evolution of key stellar quantities. The panel (a) of Figure~\ref{fig:general_evolution} displays the evolution of the accretion rate (see Equation~(\ref{eq:mdot_acc})). The stellar radius evolves in response to the accretion and the Kelvin-Helmholtz contraction as shown in the panel~(b).
We also plot a line of $R_\ast \propto t_{\rm age}^{-1/3}$ as a reference, which is predicted for the star on the Hayashi track.
This scaling is particularly relevant to the period of $t_{\rm age}\gtrsim 0.5-1$~Myr in our model.
Before the Hayashi phase, the stellar radius slightly increases due to deuterium burning in 0.1--0.3\,Myr.
The panel~(b) also shows the evolution of $r_{\rm mag}$. In the case of $B_\ast = 1$~kG, $r_{\rm mag}\approx 20~R_\odot$ at $t_{\rm age}=0.3$~Myr. As $R_\ast$ and $r_{\rm mag}$ decline similarly approximately during 0.3 to 1~Myr, we can approximate $r_{\rm mag}/R_*$ as a constant in the period. The detailed functional form is shown in Appendix~\ref{app:starevol}.
After approximately 1~Myr, the magnetospheric radius declines more slowly than the stellar radius because of reduction of the accretion rate (Equation~(\ref{eq:rmag})). 
The stellar moment of inertia $I_*$ also decreases in response to the reduction in the stellar radius (panel (c)). In the plot, $I_*$ is normalized by $I_0 = M_\odot R_\odot^2$. $k_*^2$ is the moment of inertia normalized as follows: $k_*^2 = I_* / M_* R_*^2$. 
The figure shows that $k_*^2$ is nearly constant in the period of interest, suggesting that $I_* \propto M_* R_*^2$.

The panels (d) to (f) of Figure~\ref{fig:general_evolution} display the evolution of the quantities of the stellar spin in the maximum spin rate and in the hypothetical spin equilibrium where $r_{\rm cor}=r_{\rm mag}$ is assumed. The panel (d) shows the stellar rotation periods in the two states. The period of $\sim 4$~days in the state with $r_{\rm cor}=r_{\rm mag}$ is similar to observed values for T Tauri stars \citep{Bouvier2014}. The corresponding stellar angular momenta are indicated in the panel (e). 
We write the maximum stellar angular momentum as $J_{\rm *,max}=I_*\Omega_{\rm max}$ and the angular momentum of the star in the spin-equilibrium as $J_{\rm *,eq}=f_{\rm cor,eq}^{-3/2}I_* \Omega_{\rm K}(r_{\rm mag})$.
In this study, we focus on young fast rotators and thus assume that the core and envelope rotate at the same velocity \cite[e.g.,][]{Gallet2013}.
Figure~\ref{fig:general_evolution} (f) shows $\dot{J}_{\rm CDW}$ calculated using Equation (\ref{eq:torque_CDW}). $\dot{J}_{\rm CDW}$ declines as time proceeds because the mass loss rate $\dot{M}_{\rm CDW}$ and the rotating arm $f_A r_{\rm mag}$ decrease. Therefore, we expect a strong angular momentum loss in the early phase, as argued by \citet{Ferreira2000MNRAS}.

\begin{figure}
\centering
\includegraphics[width=\columnwidth]{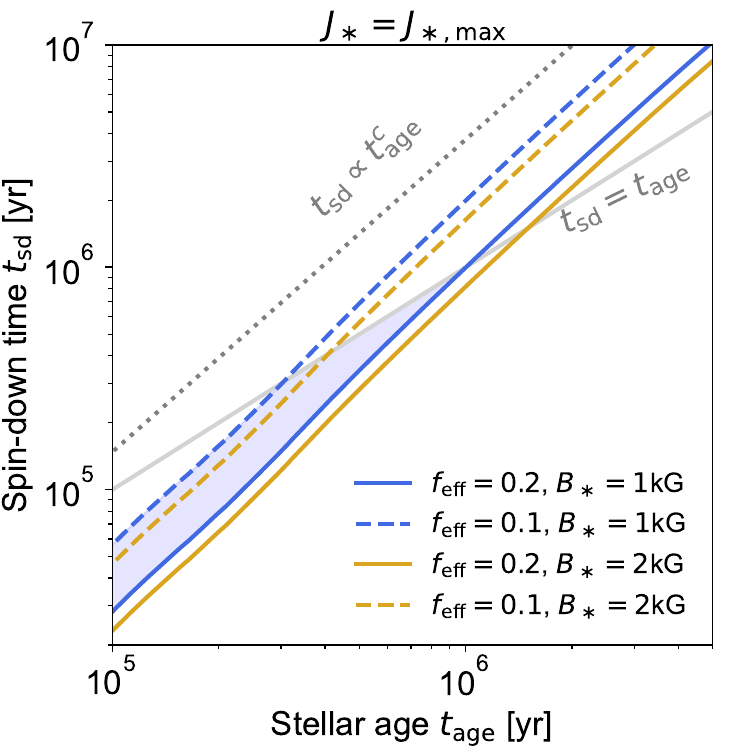}
\caption{The spin-down time $t_{\rm sd}$ for the star rotating at $\Omega_\ast=0.5\Omega_{\rm br}$ (therefore, $t_{\rm sd, up}$).
The fiducial model is indicated by the solid blue line. The region where the spin-down time for the models with the fiducial field strength (1~kG) is shorter than the stellar age is filled with blue. The gray solid line shows the line for $t_{\rm sd}=t_{\rm age}$. The dotted line shows a scaling of $t_{\rm sd}\propto t_{\rm age}^c$, where $c=59/42\approx 1.4$.}\label{fig:spindown_time}
\end{figure}

We calculate the spin-down time $t_{\rm sd}$ for the case that the stellar angular momentum is extracted only by the conical disk wind emanating from the magnetospheric boundary. $t_{\rm sd}$ is estimated as 
\begin{align}
t_{\rm sd}=\frac{J_{\ast}}{\dot{J}_{\rm CDW}}.
\end{align}
The upper limit of the spin-down time corresponds to the case that the stellar angular momentum is $J_{\ast}=J_{\rm \ast,max}$. 
We denote the upper limit as $t_{\rm sd,up}=J_{\rm \ast,max}/\dot{J}_{\rm CDW}$.
If other spin-down mechanisms such as massive stellar winds are important, the spin-down time will be smaller.
Here, we ignore the contribution of the spin-up torque by accretion. As we will see later, this assumption seems to be valid as long as the spin-up torque is significantly reduced by failed magnetospheric winds as seen in the 3D simulation (Section~\ref{subsec:overview_ST22}).

The upper limit of the spin-down time, $t_{\rm sd,up}$, is shown in Figure~\ref{fig:spindown_time}. For the fiducial case of $f_{\rm eff}=0.2$ and $B_\ast=1~$kG, the spin-down time is smaller than the stellar age in the range of $t_{\rm age}\lesssim 1$~Myr.
The result demonstrates that the conical disk wind can significantly slow down the protostar before $t_{\rm age}=1$~Myr. The result only weakly depends on the stellar field strength.  
The efficiency of the conical disk winds ($f_{\rm eff}$) has a stronger impact on the spin-down time (compare the dashed and solid lines). 
Our estimation presented here corresponds to an update of the previous estimation by \citet{Koenigl1991}.
We also derive the scaling relations of $t_{\rm sd,up}$ at the pre-MS stage, which is presented in Appendix~\ref{app:scaling_spin_down_time}.

\begin{figure}
\centering
\includegraphics[width=\columnwidth]{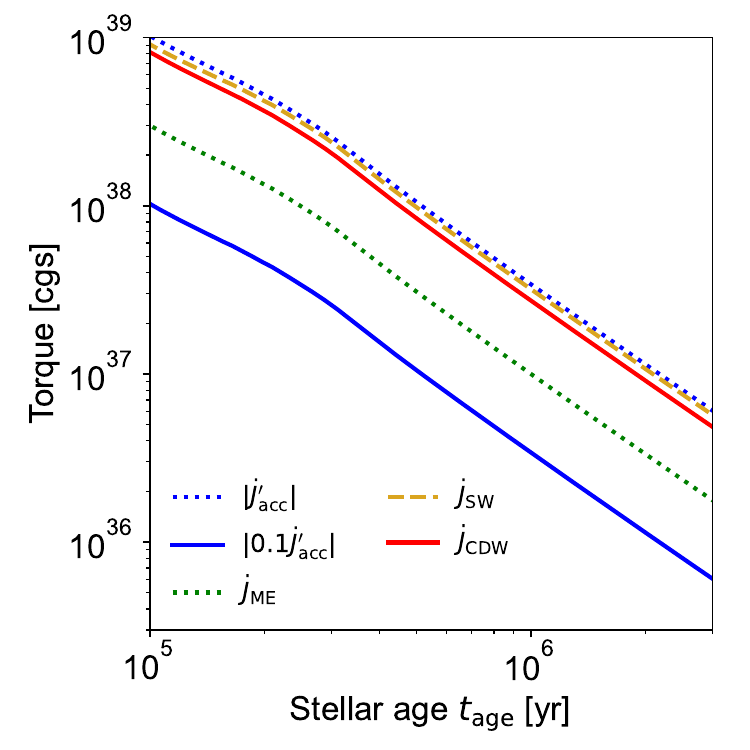}
\caption{Different torques exerting on the star with $r_{\rm cor}=0.8r_{\rm mag}$ (a reference state of the propeller regime). The red line shows the torques by the conical disk wind. 
The blue dotted line indicates the accretion torque based on the simple estimation, $\dot{J}_{\rm acc}'$. The blue solid line denotes $0.1\dot{J}_{\rm acc}'$, an accretion torque that takes into account the angular momentum extraction by some winds.
Other torques are also plotted for comparison.
The green dotted line shows the spin-down torque by magnetospheric ejections. The dashed golden line denotes the spin-down torque by the stellar wind in the case that $\dot{M}_{\rm SW}=0.1\dot{M}_{\rm acc}$.
}\label{fig:torque}
\end{figure}

To illustrate the significance of the spin-down torque by the conical disk winds, we compare it with other possible spin-down torques discussed by previous studies:
\begin{itemize}
    \item the torque by the stellar wind, $\dot{J}_{\rm SW}$
    \item the torque by the magnetospheric ejection, $\dot{J}_{\rm ME}$
\end{itemize}
As they depend on the stellar spin rate, a direct comparison requires a detailed calculation of the time evolution of the stellar spin rate. As this is beyond the scope of this study, we calculate these torques in a propeller regime by assuming that $r_{\rm cor}=0.8r_{\rm mag}$.

We briefly describe the calculation methods for the above two torques. We provide the detailed expressions for $\dot{J}_{\rm SW}$ and $\dot{J}_{\rm ME}$ in Appendix~\ref{app:torques_ME_SW}.
For the mass loss rate of the stellar wind, $\dot{M}_{\rm SW}$, we assume $\dot{M}_{\rm SW}=f_{\rm SW}\dot{M}_{\rm acc}$ and $f_{\rm SW}=0.1$, which would be a minimum value for the stellar wind to spin down the protostar \citep[e.g.][]{Gallet2019,Ireland2021}. We note that this high efficiency ($\sim 10$\%) seems to be difficult to realize in reality \citep{Cranmer2008,Zanni2011}.
For $\dot{J}_{\rm ME}$, we adopt the analytical expression by \citet{Gallet2019} based on the axisymmetric models \citep{Livio1992MNRAS,Armitage1996MNRAS,Matt2005}. We note that the magnitude of the torque could be highly overestimated because of the axisymmetric assumption (see Section~\ref{subsec:overview_ST22}).
Therefore, we should take the value as a reference.

Figure~\ref{fig:torque} compares different torques which can exert on the star in the propeller regime with $r_{\rm cor}=0.8r_{\rm mag}$. 
To calculate $\dot{J}_{\rm CDW}$, we use our Equation~(\ref{eq:torque_CDW}) and adopt $f_{\rm eff}=0.2$ and $B_\ast=1$~kG. Namely, we assume that $\dot{J}_{\rm CDW}$ does not explicitly depend on the stellar spin rate $\Omega_{\rm ast}$.
The accretion torque which takes into account the 3D effect, $\dot{J}_{\rm acc}=0.1\dot{J}_{\rm acc}'$, is also shown.
$\dot{J}_{\rm CDW}$ is similar to $\dot{J}_{\rm SW}$ for the given parameter set, which indicates that the conical disk winds can provide as a large spin-down torque as the hypothetical massive stellar winds.
If the actual accretion torque is $\sim 0.1 \dot{J}_{\rm acc}'$ or smaller as suggested by the 3D MHD simulation, the spin-down torques can dominate the spin-up torque.
The spin-down torque by magnetospheric ejections is unimportant in this case, but the torque is sensitive to the choice of $r_{\rm cor}$. If we adopt $r_{\rm cor}=0.5r_{\rm mag}$, we find that $\dot{J}_{\rm ME} \sim \dot{J}_{\rm CDW}$. Again, we note that the magnitude is likely to be overestimated as the formula of $\dot{J}_{\rm ME}$ is based on axisymmetric models. If magnetospheric ejections are indeed the primary mechanism for angular momentum transport, we require a theory to account for the observed weak time variability in protostellar accretion.

\section{Discussion}
Considering numerical results of \citet{Takasao2022} (ST22), we propose that the magnetically driven winds just around the magnetosphere are key to resolving the spin-down problem. Failed magnetospheric winds, which only appear in three-dimension, significantly reduce the spin-up torque. When the protostar is in the propeller regime, the powerful conical disk wind will appear and extract the stellar angular momentum. A combination of the two leads to an efficient angular momentum loss.
The mass loss rate can be $\sim 10$\% of the accretion rate as a result of direct mass loading from the inner disk to the rotating magnetosphere (Appendix~\ref{app:mass_loss_rate}). 
Our study shows that a higher accretion rate in the earlier phase leads to a larger wind mass loss rate. This is the main reason why the younger protostars have larger spin-down torque by the conical disk wind.

Recent models of spin evolution hypothesize the presence of massive stellar winds \cite[e.g.,][]{Gallet2019, Gehrig2022}, but this assumption has been challenged from an energetic perspective. 
Our estimation suggests that, in the propeller regime, the conical disk wind will play a critical role in stellar angular momentum loss.
If stellar winds play a significant role, the spin-down mechanism will depend heavily on the properties of the accreting object. The detailed spin evolution is influenced by the initial conditions, stellar evolution (e.g., stellar contraction), and disk evolution. Thus, to reach a more robust conclusion, it is essential to incorporate all key factors into a comprehensive model.

In Section~\ref{subsec:overview_ST22}, we have argued that the interaction between the stellar fields and disk fields determines how the mass and angular momentum transfer (see also Appendices~\ref{app:angular_mom_flows} and \ref{app:mass_loss_rate}). Here, we briefly note the importance of magnetic reconnection between the stellar and disk poloidal fields, which is not investigated in details in this study. \citet{Ferreira2000MNRAS} argued that efficient mass loading via magnetic reconnection to the rotating winds will be key in increasing their spin-down torque. On the other hand, \citet{Romanova2011MNRAS} performed 2D MHD simulations and showed that magnetic reconnection reduces the accretion torque by decreasing the total magnetic flux of the stellar fields threading the disk \citep[see also][]{Parfrey2017ApJ}.
How the disk fields affect the stellar spin evolution in 3D will be an interesting topic for future studies.
As our model \citep{Takasao2022} is initialized with a magnetized torus, magnetic reconnection between the stellar and disk poloidal fields will be operating (the presence of the disk fields is a noticeable difference from \citet{Zhu2024MNRAS}). However, the plasma $\beta$ based only on the poloidal fields is much larger than unity in the inner disk outside the magnetosphere, which suggests a minor role of such reconnection in our model.

There are some caveats about our numerical simulations. The 3D simulations of ST22 which motivate this theoretical study have smaller magnetospheric radii ($\sim 2R_\ast$) than the value used in this study, as we adopted a relatively weak stellar field strength ($\sim 160~{\rm G}$) to avoid numerical difficulties. The relation between $r_{\rm m}$ and $r_A$ for more realistic situations is to be studied. Another issue is the numerical treatment of the stellar wind. Since the properties of stellar winds remain unclear observationally, it is important to investigate how the results of this study depend on the stellar wind model. A detailed consideration on the efficiency of the conical disk wind $f_{\rm eff}$ is also necessary. The dependence on the stellar spin will particularly affect the behavior of the star near the spin-equilibrium. As the efficiency is relevant to turbulent mixing, convergence check with higher spatial resolution simulations is also a remaining task.

Studying stellar spin is crucial for testing scenarios of stellar and disk evolution. 
Observations have revealed the diversity in the structure of protoplanetary disks \citep{Bae2022}. The difference in the disk structure will lead to different accretion histories such as episodic accretion outbursts. As our study indicates the tight relationship between accretion and spin-down, it is possible that the stellar spin distribution is a consequence of the diversity in the disk accretion histories \citep[see also][]{Gehrig2023}.
Stellar spin also affects chemical mixing in the stellar interior and thus stellar evolution.
Lithium may be a good tracer of the history of internal mixing because its abundance is sensitive to it.
Some studies suggest that models which take into account the effects of rotational mixing may explain the origin of the lithium-depleted stars \citep{Bouvier2016,Eggenberger2022}. 
Accurate interpretation of the stellar surface abundance is important for testing planet formation scenarios because the surface abundance would depend not only on the internal mixing but also on how planet formation affects the abundance of accreting materials \citep{Kunitomo2022}. Further studies of spin evolution are thus required for our understanding of how star-planet systems including the solar system are born and evolve.

We thank the anonymous referee for his/her careful reading and insightful comments.
S.T. was supported by JSPS KAKENHI grant No. JP22K14074. M.K. was supported by the JSPS KAKENHI grant Nos. JP20K14542, JP23H01227, JP24K07099, and 24K00654, and thanks Observatoire de la Côte d’Azur for the hospitality during his long-term stay in Nice. S.T., K.I. and K.T. were supported by the JSPS KAKENHI grant Nos. JP21H04487 and JP22KK0043. T.K.S. was supported by the JSPS KAKENHI No. JP22H01263.
Numerical computations were carried out on Cray XC50 and PC cluster at the Center for Computational Astrophysics, National Astronomical Observatory of Japan. Test calculations in this work were in part carried out at the Yukawa Institute Computer Facility. This work was supported by MEXT as a Program for Promoting Researches on the Supercomputer Fugaku by the RIKEN Center for Computational Science ``Toward a uniﬁed view of the universe: from large-scale structures to planets" (Grant Number JPMXP1020200109) and ``Structure and Evolution of the Universe Unraveled by Fusion of Simulation and AI" (Grant Number JPMXP1020230406).

\appendix
\section{Analysis of angular momentum flux}\label{app:angular_mom_flows}

\begin{figure*}
    \centering
    \includegraphics[width=1.0\columnwidth]{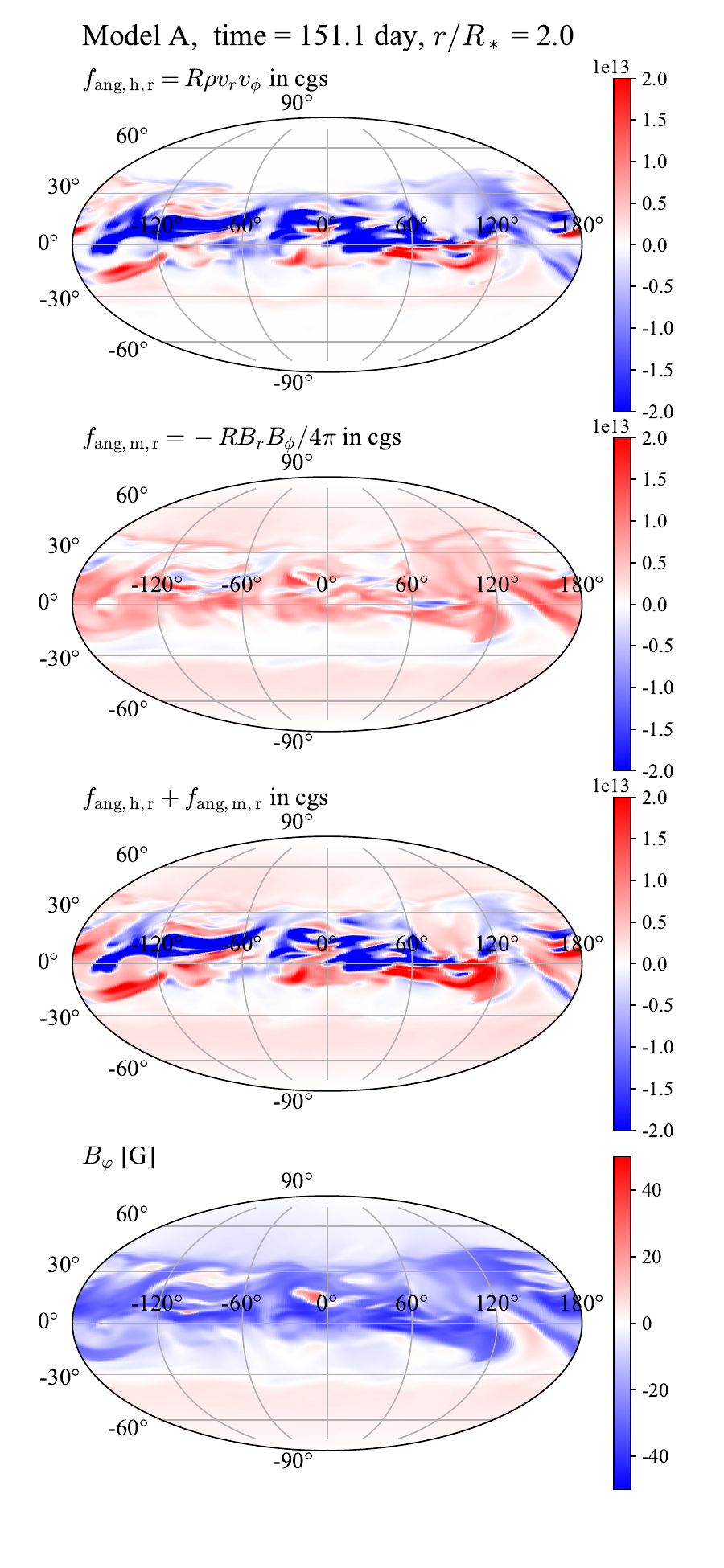}
    \includegraphics[width=1.0\columnwidth]{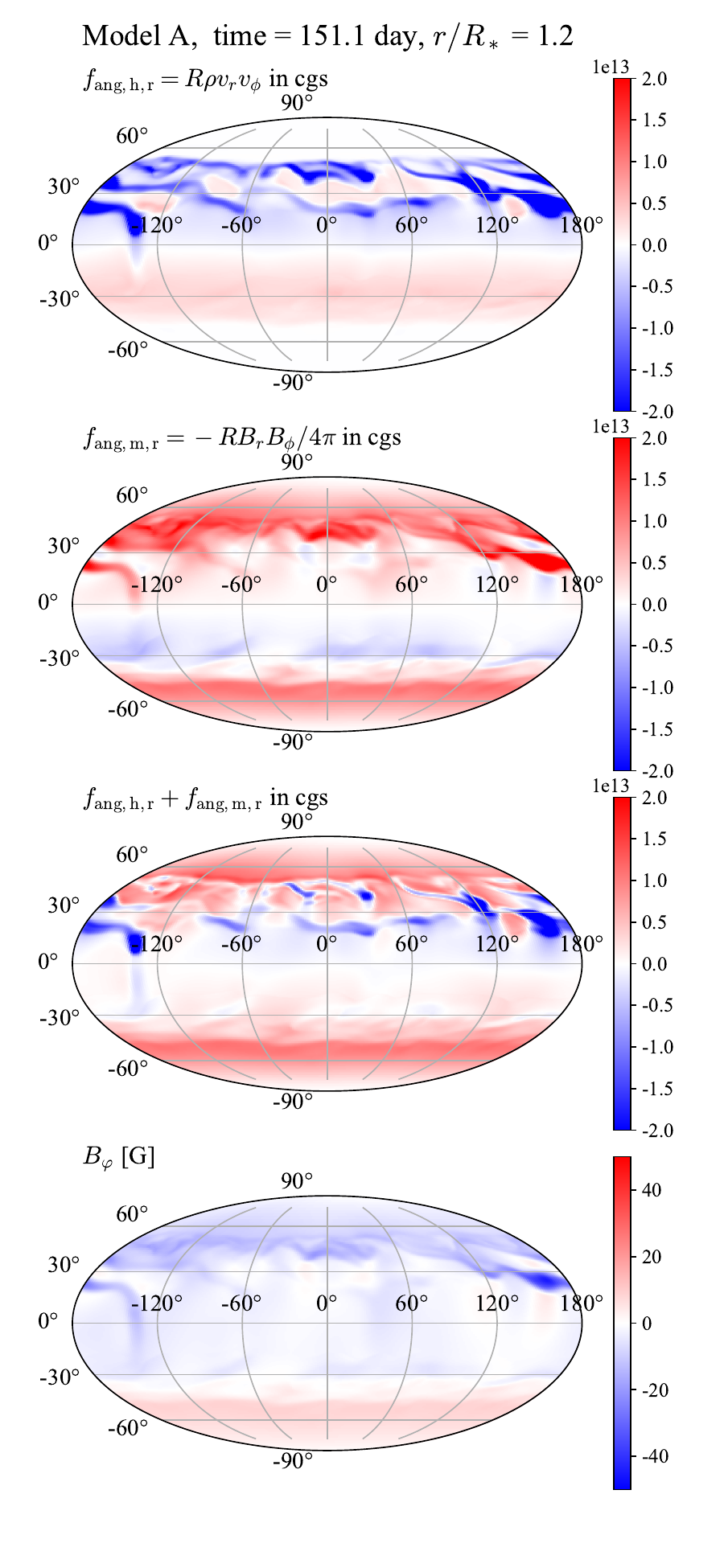}
    \caption{Analysis of angular momentum transfer at $r=2.0R_\ast$ (left) and $1.2R_\ast$ (right) for Model~A of ST22. From top to bottom, $f_{{\rm ang, h},r}=R\rho v_r v_\varphi$, $f_{{\rm ang, m},r}=-RB_r B_\varphi/4\pi$, $f_{{\rm ang, h},r}+f_{{\rm ang, m},r}$, and $B_\varphi$. Note that accretion mainly occurs in the northern hemisphere.}
    \label{fig:ang_mom_mlwd}
\end{figure*}

We investigate the angular momentum flow around the protostar by analyzing the spatial distributions of the angular momentum flux and toroidal fields on some spherical surfaces. The left column of Figure~\ref{fig:ang_mom_mlwd} displays the result at the spherical radius of $r=2R_\ast$, which is close to the magnetospheric radius. 
The Reynolds stress (the top panel) takes a large negative value around the midplane, which means that accreting flows are transporting angular momentum inward. We also see some patchy regions with positive values. Some of them correspond to failed magnetospheric winds (see Section~\ref{subsec:overview_ST22}). The Maxwell stress shows positive values in almost all the directions (the second panel). It takes larger positive values around the midplane because strong toroidal fields are there in response to the magnetosphere-disk interaction (the bottom panel). The sum of the Reynolds and Maxwell stress results in the spin-down torque as a net (ST22). We note that the total flux displays a complicated structure which is highly nonuniform in the azimuthal direction (the third panel), which is distinct from the picture based on the axisymmetric model.

The right column of Figure~\ref{fig:ang_mom_mlwd} shows the result near the stellar surface, $r=1.2R_\ast$. The Reynolds stress displays a spotty structure which is formed by patchy accretion flows. The Reynolds stress shows negative values only in the northern hemisphere as asymmetric accretion occurs. The Maxwell stress shows large positive values in the range of $30^\circ \lesssim |\theta| \lesssim 60^\circ$. The spin-down Maxwell stress is produced by the back-reaction of driving the conical disk winds and failed magnetospheric winds. Indeed, a large part of the stellar field lines driving these winds emanate from the latitudinal range (Figures~\ref{fig:3D} and \ref{fig:penetrating_flow}). The total flux indicates that a large fraction of the spin-up torque by accreting flows is compensated by the magnetic spin-down torque, which results in a smaller rate of the angular momentum injection than the classical estimation, $\dot{J}_{\rm acc}^\prime$.

\section{Structure of the Alfv\'en surface}\label{app:Alfven_surface}
\begin{figure}
\centering
\includegraphics[width=\columnwidth]{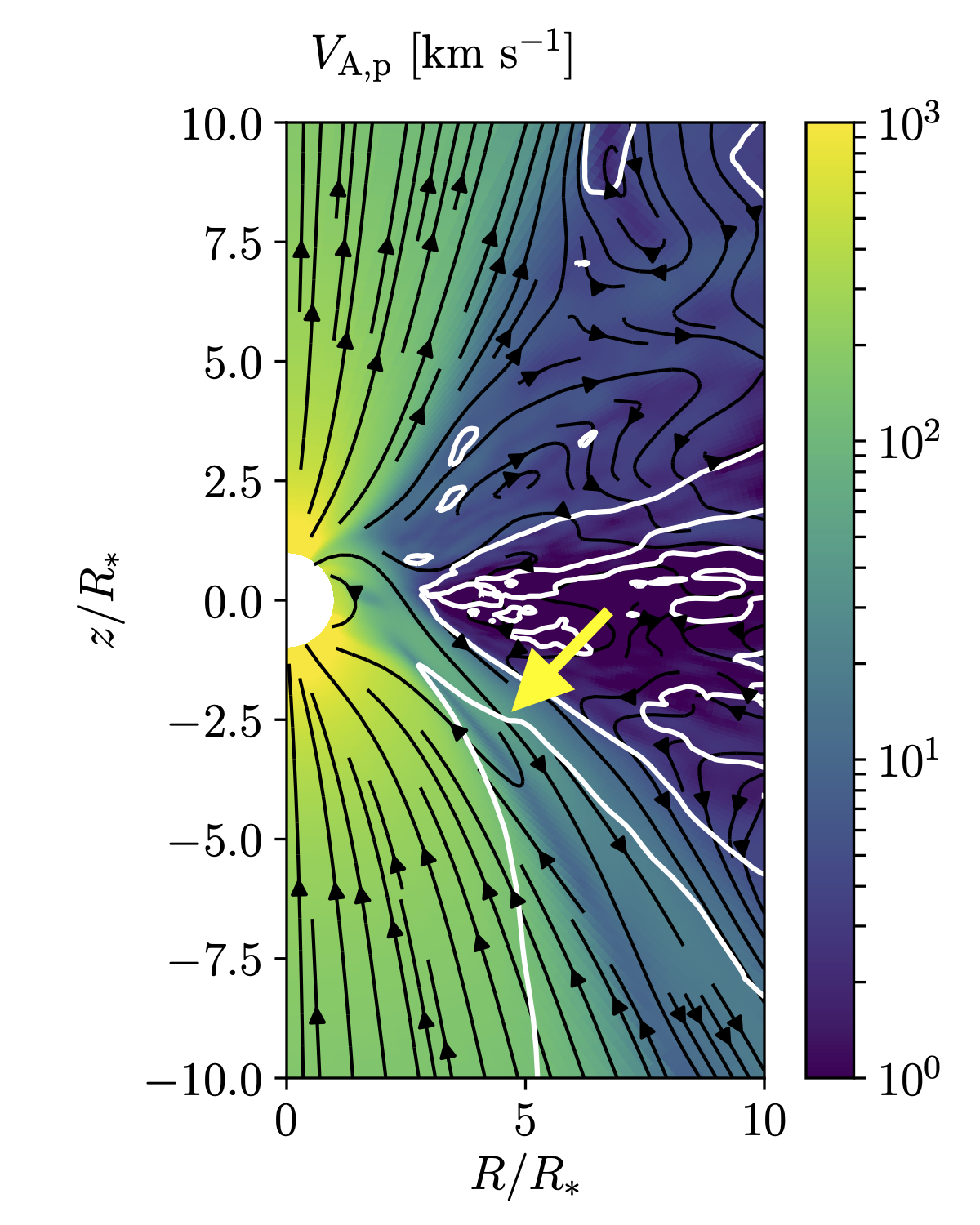}
\caption{The color denotes the poloidal Alfv\'en speed. The white contours indicate Alfv\'en surfaces where the poloidal velocity is equal to the poloidal Alfv\'en speed. Black lines with arrows denote magnetic field lines projected on this plane. The yellow arrow indicates the location of the Alfv\'en point discussed in the text of Section~\ref{app:Alfven_surface}. The data are temporally and azimuthally averaged. The time average is performed during the period of $t=190.1–199.4$ days after the simulation starts.}\label{fig:Alfven_radius}
\end{figure}

Figure~\ref{fig:Alfven_radius} shows the structure of the Alfv\'en surface. The color denotes the poloidal Alfv\'en speed $V_{\rm A,p}$, and the black lines with arrows indicate the averaged poloidal field structure. The magnetosphere expands in the southern hemisphere, and the conical disk wind is blowing along the expanding field lines. The white lines indicate the locations where the poloidal plasma velocity is equal to $V_{\rm A,p}$. The white line in the conical disk wind region shows the Alfv\'en surface.

ST22 demonstrates that the outward angular momentum flux takes the largest value around at the latitude of approximately $45^{\circ}$ or slightly larger at the stellar surface (see the right column of Figure~\ref{fig:ang_mom_mlwd} of this study and Figure~19 of ST22). 
Note that the field lines driving the conical disk winds are connected with the northern hemisphere of the star.
Considering this result, we focus on the field line emanating from the latitude of approximately $45^{\circ}$ at the stellar surface. The field line intersects the Alfv\'en surface approximately at the cylindrical radius of $5R_\ast$ (see the location indicated by the yellow arrow). As the magnetospheric radius is approximately $2.5R_\ast$, we find $r_A\approx 2r_{\rm mag}$. Therefore, we adopt $f_A=2$ as a fiducial value in this study. A detailed dependence of the Alfv\'en radius on the properties of accretion and stellar magnetic fields should be investigated in future studies. The Alfv\'en surface inside the electric current sheet of the expanding magnetosphere comes closer to the protostar ($\sim 2R_\ast$). However, because the field strength in the current sheet is weaker than the surrounding, the angular momentum transport inside the current sheet is unimportant.

\section{Theoretical estimation of mass loss rate of conical disk wind}\label{app:mass_loss_rate}
\begin{figure}
\centering
\includegraphics[width=\columnwidth]{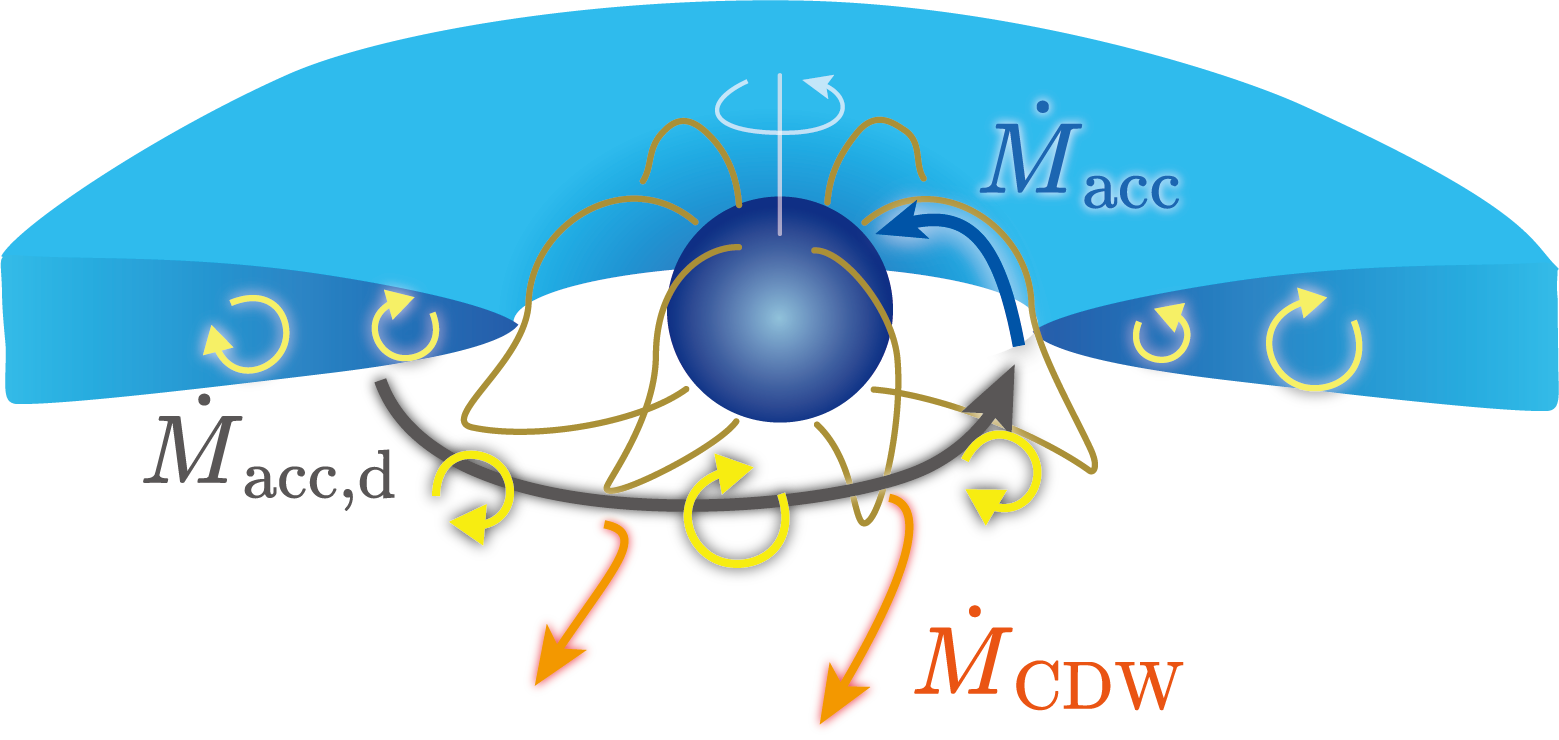}
\caption{Illustration of the bifurcation of the mass flow. The black arrow indicates the accreting flow in the disk. A fraction of the accreting gas is loaded to the rotating stellar magnetic fields via turbulence (indicated as yellow arrows) and becomes the conical disk wind (orange arrows). The rest of it falls onto the star (dark blue arrow). }\label{fig:mass_bifurcation}
\end{figure}

Let us define the accretion rate in the disk as $\dot{M}_{\rm acc,d}$. From the law of conservation of mass, we get
\begin{align}
    \dot{M}_{\rm acc,d}= \dot{M}_{\rm acc}+\dot{M}_{\rm CDW}.\label{eq:mass_cons}
\end{align}
The majority of the accreting material falls onto the protostar as a result of angular momentum loss ($\dot{M}_{\rm acc}$). The rest of it is loaded onto the rotating magnetospheric field via mixing and is ejected away as the conical disk wind ($\dot{M}_{\rm CDW}$). Figure~\ref{fig:mass_bifurcation} displays a schematic illustration of the bifurcation of the mass flow. Here, we investigate the bifurcation ratio.

We first note the difference in magnetosphere-disk interaction between 2D and 3D models. In 2D models, the accreting material enters the magnetosphere as a result of effective turbulent diffusion (see also the left panel of Figure~\ref{fig:summary_acc_ejec}). In other words, accreting material penetrates the magnetosphere as a result of diffusive mixing (in such models, the bifurcation ratio is sensitive to the assumed effective diffusivity and viscosity \citep{Ustyugova2006}, which are highly uncertain).
In three-dimension, the mass loading to the magnetosphere occurs differently. A part of the disk mass is loaded through the turbulent mixing at the magnetospheric boundary, which may be modeled using an effective diffusivity as done in the 2D models. In addition, the 3D model shows filamentary flows penetrating into the magnetosphere (Figures~\ref{fig:penetrating_flow} and \ref{fig:summary_failed_winds}). The penetrating flows posses strong toroidal fields because they are dragging the disk toroidal fields and continuously shearing them up (see Section~\ref{subsec:overview_ST22}). They retain their coherent structure even in the magnetosphere possibly because their strong toroidal fields prevent the flows from breaking up.
Their coherent structure motivates us to treat the mixing and penetration of the accretion flows separately. The mixing operates mainly at the magnetospheric boundary, which is outside the corotation radius in the propeller regime. Therefore, we expect that most of the gas loaded via the mixing will be blown away by a combination of the centrifugal and Lorentz forces \citep{Ustyugova2006,Romanova2009MNRAS}. However, the penetrating flows can go inside the magnetosphere due to their large inertia.

The bifurcation ratio of the mass flow will depend on the rates of gas penetration and mixing.
The rate for the accreting material in the magnetosphere to fall onto the protostar is expressed as $t_{\rm acc,mag}^{-1}$. The rate for mixing (which is essential for mass loading to the conical disk wind) is written as $t_{\rm mix}^{-1}$. 
By using them, we can write the rate of mass loading to the magnetosphere, $\dot{M}_{\rm load}$, as follows:
\begin{align}
    \dot{M}_{\rm load} &= \frac{t_{\rm mix}^{-1}}{t_{\rm mix}^{-1}+t_{\rm acc,mag}^{-1}}\dot{M}_{\rm acc,d} \nonumber \\
    &=\frac{t_{\rm acc,mag}}{t_{\rm mix}+t_{\rm acc,mag}}\dot{M}_{\rm acc,d}.
\end{align}
A fraction of the loaded mass will be blown away as the conical disk wind, while the rest of it will accrete onto the protostar as a funnel flow. 
By introducing a nondimensional parameter $f_{\rm CDW}$, which denotes the partition rate, we obtain
\begin{align}
    \dot{M}_{\rm CDW}=f_{\rm CDW}\dot{M}_{\rm load}.\nonumber
\end{align}
The value of $f_{\rm CDW}$ should depend on the details of the dynamics (e.g. the distributions of the Lorentz and thermal forces along the field line) and the geometrical effects (e.g. the north-south asymmetry). Nevertheless, we expect that a large portion of the loaded mass will gain the angular momentum from the protostar as the magnetospheric boundary is outside the corotation radius: $r_{\rm cor}<r_{\rm mag}$ (therefore, we expect that $f_{\rm CDW}$ is comparable to unity).
This assumption should be examined in future studies. From the law of conservation of mass (Equation~\ref{eq:mass_cons}), we get
\begin{align}
    \dot{M}_{\rm CDW} = \frac{f_{\rm CDW}t_{\rm acc,mag}}{t_{\rm mix}+(1-f_{\rm CDW})t_{\rm acc,mag}}\dot{M}_{\rm acc}
\end{align}
If $f_{\rm CDW}$ is insensitive to the dynamics and geometrical effects at the magnetospheric boundary, this equation explains why $\dot{M}_{\rm CDW}$ is proportional to $\dot{M}_{\rm acc}$.

We estimate $t_{\rm acc,mag}$ and $t_{\rm mix}$ by referring to the results of ST22. As the gas penetrating the magnetosphere falls onto the protostar within one to two orbital rotation periods, we take $t_{\rm acc,mag}\approx t_{\rm K}(r_{\rm mag})$, where $t_{\rm K}(r)$ denotes the Keplerian orbital time at the radius of $r$. The timescale of mixing should be related to the level of the velocity fluctuation (or root-mean-square fluctuating velocity) around the magnetospheric boundary, $\delta v$. 
The diffusion coefficient due to turbulence in the Keplerian disk can be expressed as 
\begin{align}
    D \approx \frac{1}{3}\frac{\langle \delta v^2\rangle}{\Omega_{\rm K}},
\end{align}
where $\langle \delta v^2\rangle$ denotes the azimuthally and temporally averaged turbulent velocity. Using this expression, the mixing timescale can be estimated as
\begin{align}
    t_{\rm mix}=\frac{\Delta R^2}{D},
\end{align}
where $\Delta R$ is the width of the transition layer between the magnetosphere and the disk.

\begin{figure}
\centering
\includegraphics[width=\columnwidth]{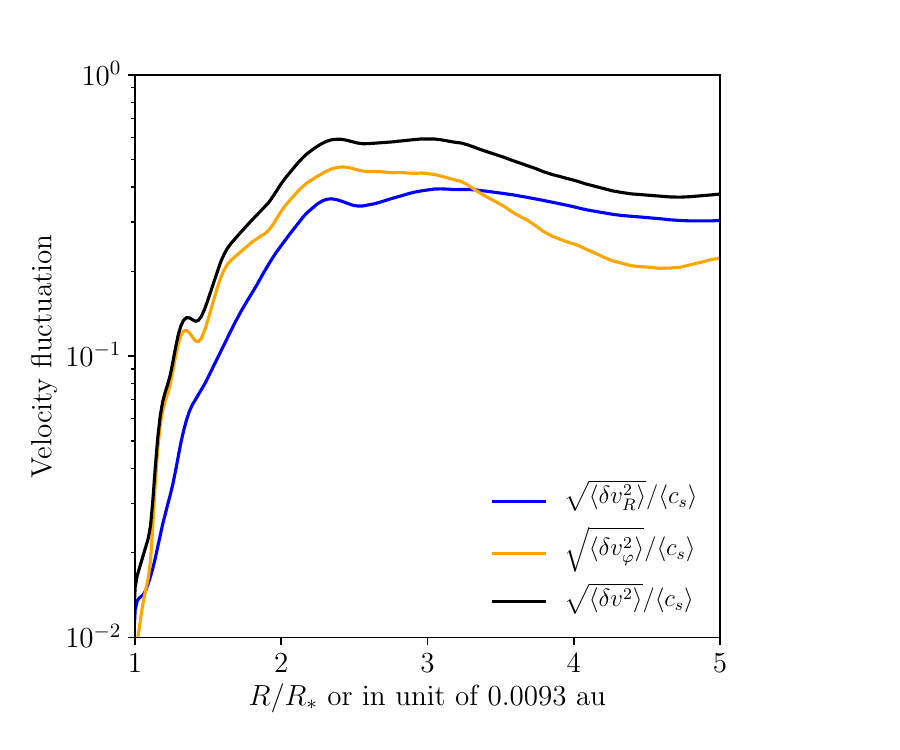}
\caption{The velocity fluctuation measured around the equatorial plane. The time average is performed during the period of $t=190.1–199.4$ days after the simulation starts.}\label{fig:vel_fluctuation}
\end{figure}

We measure the velocity fluctuation in the simulation. Figure~\ref{fig:vel_fluctuation} displays the velocity fluctuation measured around the equatorial plane. The bracket indicates the azimuthally and temporally averaged quantities. $\delta v_R$ and $\delta v_\varphi$ are the radial and azimuthal components, respectively, which are the most relevant to the mixing. $c_s$ is the sound speed. We calculate $\delta v$ as $\delta v^2=\delta v_R^2+\delta v_\varphi^2$. The figure indicates that $\sqrt{\langle\delta v^2\rangle}/\langle c_s\rangle \approx 0.5$ around the magnetospheric boundary ($R\approx 2.5R_\ast$).
We note that it is likely that the value presented here depends on the numerical resolution. The convergence check is the remaining task. Nevertheless, we take the measured value as a fiducial value for a better comparison between our simulation and order-of-magnitude calculation.
As the vortex size is expected to be limited by the disk thickness or the pressure scale height $H$, we can rewrite $t_{\rm mix}$ as
\begin{align}
    t_{\rm mix}\approx 4 t_{\rm K}(r_{\rm mag})\left( \frac{\Delta R}{H}\right)^{2}\left( \frac{\sqrt{\langle \delta v^2\rangle}/\langle c_s\rangle}{0.5}\right)^{-2},
\end{align}
where we have used the relation $H=\sqrt{2}c_{\rm s}/\Omega_{\rm K}$.

Using the above parameter sets, we finally find
\begin{align}
    \frac{\dot{M}_{\rm CDW}}{\dot{M}_{\rm acc}}=f_{\rm CDW}\frac{t_{\rm acc,mag}}{t_{\rm mix}}\approx 0.2 \left(\frac{f_{\rm CDW}}{0.8}\right),
\end{align}
where we set the order-unity parameter $f_{\rm CDW}$ to be $0.8$ by considering imperfect mass loading to the conical disk winds.
This result indicates that $f_{\rm eff}=0.1-0.2$ is a reasonable value. A larger $f_{\rm eff}$ may be possible if the velocity fluctuation level is larger than the one considered here \citep{Ustyugova2006}. 
Rapid rotators may show a larger fluctuation level because the toroidal fields at the magnetospheric boundary are more strongly amplified than slow rotators (Figure~11 of ST22).
We note that the discussion here only focuses on the bifurcation ratio near the wind launching region. The above mass loss rate should be larger than the mass loss rate of the gas escaping from the stellar gravity because not all the wind gas will escape from the system. If the acceleration is insufficient, a fraction of the wind gas will fall back to the disk or the protostar. It is worth noting that the above estimate is close to but a factor of a few larger than the typical ratio of the mass outflow rate of the jet to the accretion rate \citep[0.05--0.1. See, e.g.,][]{Fang2018}.
We also note that the above discussion ignores the role of magnetic reconnection between the stellar and disk poloidal fields \citep{Ferreira2000MNRAS}. If the reconnection plays critical role in the mass loading, we have to consider the effect.

\section{Comparison of mass loading between the stellar wind and the conical disk wind}\label{app:comparison_mass_loading}

\begin{figure}
\centering
\includegraphics[width=\columnwidth]{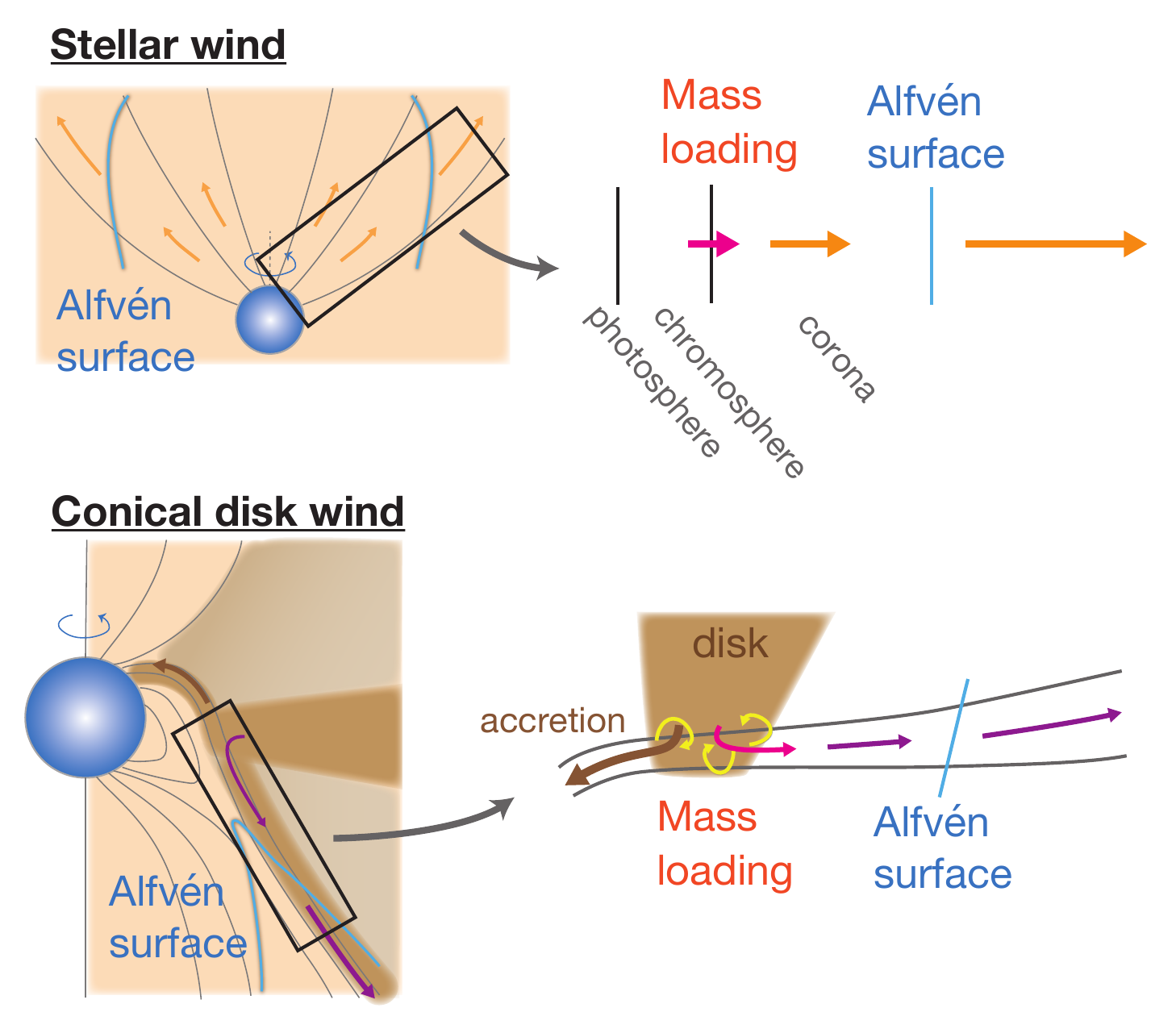}
\caption{Comparison of mass loading between the stellar wind (top) and the conical disk wind (bottom). Note that in both cases, mass loading occurs inside the Alfv\'en surfaces.}\label{fig:comp_mass_loading}
\end{figure}

Figure~\ref{fig:comp_mass_loading} compares mass loading between the stellar wind and the conical disk wind. In the case of stellar wind, the mass is loaded from the stellar atmosphere. This is also true for the accretion-powered stellar wind models \citep{Matt2005,Cranmer2008}.
The coronal gas which passes through the Alfv\'en surface becomes the stellar wind.
In the case of the conical disk wind, the wind mass is loaded from the accretion disk at a distant place from the stellar surface. Small-scale magnetic reconnection driven by turbulent motions allows accreting materials to get on the rotating stellar magnetic fields. 
In both cases, mass loading occurs inside the Alfv\'en surfaces. However, in the case of conical disk wind, the mass is loaded where the gravitational potential is shallower than at the stellar surface. Therefore, blowing the massive wind is less expensive in terms of energy.

\citet{Ferreira2000MNRAS} also notice the advantage of the conical disk wind. They studied the situation in which the disk has a poloidal field that can reconnect with the protostellar magnetic fields, which is likely in the early phase of star formation \citep[see also][]{Hirose1997PASJ}. The reconnection creates an rotating open field and loads the disk gas to the field simultaneously, which leads to the formation of the conical wind mediated by reconnection (they call it reconnection X-wind).
They showed that the protostar driving reconnection X-wind can spin-down in a way consistent with observations.

The mass loading in 3D is similar to what is assumed in \citet{Ferreira2000MNRAS} in the sense that magnetic reconnection is relevant. However, the 3D model suggests that the presence of a large-scale disk poloidal fields will not be necessary for the reconnection-mediated mass loading. The disk in our 3D model has a finite poloidal field, but the azimuthally averaged poloidal field of the disk appears to be turbulent (see Figure~6 of ST22). Therefore, the strength of the disk field is too weak for the reconnection X-wind to blow. Our study demonstrates that the reconnection-mediated conical disk wind plays important roles in the stellar spin-down in a broader situation than previously considered.

\section{Models of stellar evolution and accretion} \label{app:starevol}

We calculated the evolution of young stars following \citet{Kunitomo2021}.
We used the Modules for Experiments in Stellar Astrophysics (MESA) stellar evolution code version 12115 \citep{Paxton2011}. We refer the reader to \citet{Kunitomo2021} and the series of papers by Paxton et al. for details of the computational method in this work.

We started from a protostellar phase with a seed of mass $0.1\,M_\odot$. We adopted the accretion rate $\dot{M}=10^{-5}\,M_\odot\,\rm yr^{-1}$ for $t_{\rm age}\leq t_1$ and $\dot{M}=10^{-5}\,M_\odot\,{\rm yr^{-1}}\times(t_{\rm age}/t_1)^{-1.5}$ for $t_1<t_{\rm age}<10^7$\,yr following \citet{Hartmann1998}, where $t_1=31,160\,$yr \citep[Figure\,3 of][]{Kunitomo2021}. The resulting final mass is $1M_\odot$. We neglect the effects of rotation and stellar winds on stellar evolution.

\begin{figure*}
\centering
\includegraphics[width=2\columnwidth]{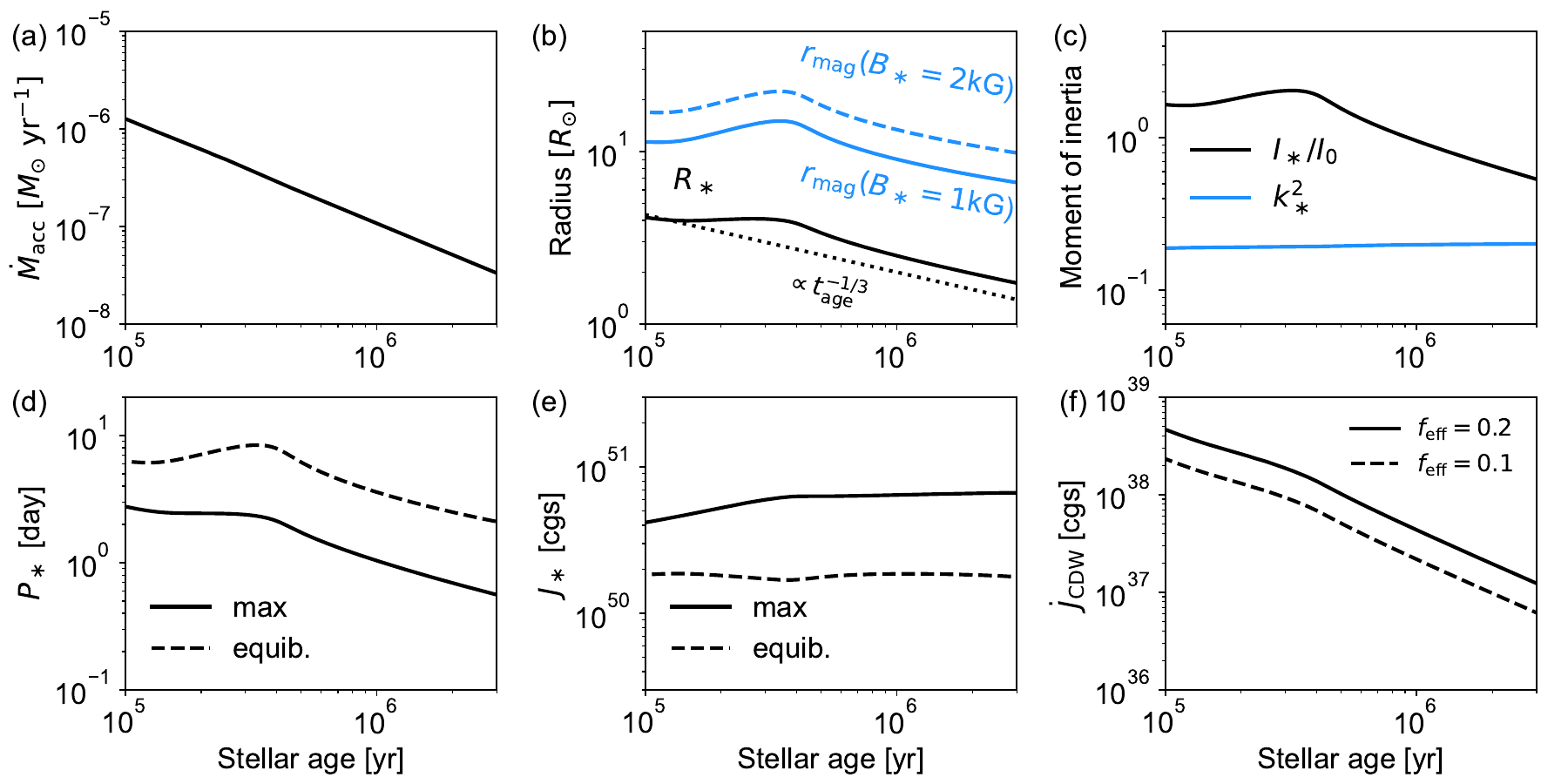}
\caption{Same as Fig.\,\ref{fig:general_evolution} but with another accretion rate \citep{Hartmann2016}.}\label{fig:general_evolution_H16}
\end{figure*}

\begin{figure}
\centering
\includegraphics[width=\columnwidth]{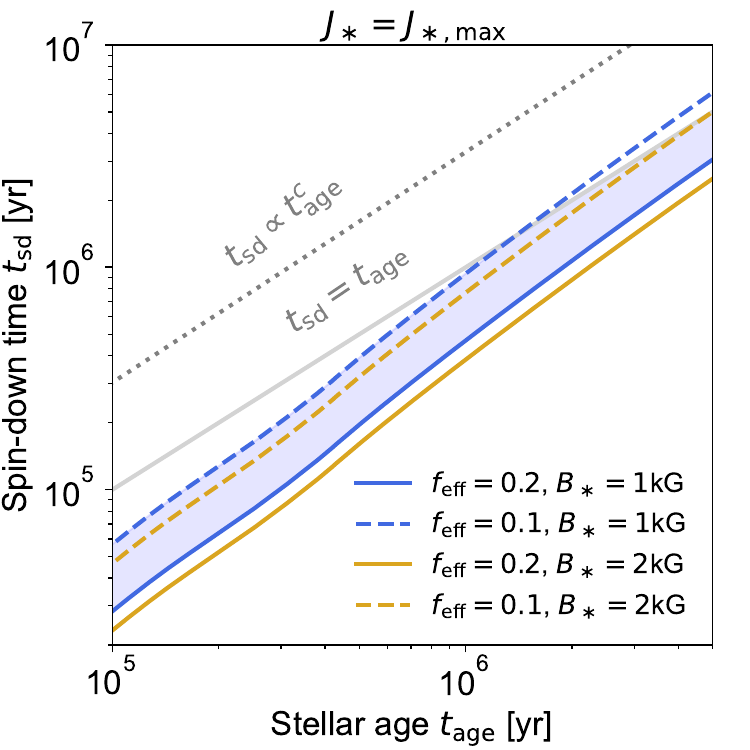}
\caption{Same as Fig.\,\ref{fig:spindown_time} but with another accretion rate \citep{Hartmann2016}.}\label{fig:spindown_time_H16}
\end{figure}

We note that the accretion rate $\dot{M}$ is still uncertain. Our fiducial model ($\dot{M} \propto t_{\rm age}^{-1.5}$) based on \citet{Hartmann1998} seems reasonable in the viewpoint of the viscous accretion with a constant viscosity $\alpha$ parameter \citep{Shakura1973} \citep[see discussions in][]{Hartmann1998}. However, recent observational studies have suggested another empirical relation in which $\dot{M}\propto t_{\rm age}^{-1.07}$ \citep{Hartmann2016}.
To see the impact of the uncertainty on the conclusion, we have simulated a protostellar evolution with the latter accretion rate. The resulting final stellar mass is $1M_\odot$, as in the fiducial model.
Figures\,\ref{fig:general_evolution_H16} and \ref{fig:spindown_time_H16} show the evolution of stellar key quantities and spin-down time, respectively, as in Figs.\,\ref{fig:general_evolution} and \ref{fig:spindown_time}. Due to the higher accretion rate in the late phase, this model results in an even shorter spin-down time than the fiducial model in Fig.\,\ref{fig:spindown_time} (see also Appendix\,\ref{app:scaling_spin_down_time}). Therefore, we conclude that the conclusion of this study (i.e., the successful spin-down of protostars due to CDW) is not affected by the uncertainties in the accretion rate.

The model used in this work is the same as the model ``K2'' in \citet[][see their Table\,1]{Kunitomo2021} except for accretion heating\footnote{The data and the \texttt{inlist} files for MESA simulations are available on Zenodo under an open-source 
Creative Commons Attribution license: 
\dataset[doi:10.5281/zenodo.14524940]{https://doi.org/10.5281/zenodo.14524940}.}. \citet[][see their Section 3.1.2]{Kunitomo2021} modeled the accretion heating with $\xi=0.1$, whereas in this study we used $\xi=0.5$ (i.e., higher-entropy accretion) resulting in the evolution more similar to the classical one. Since the $A_2$ parameter, which controls opacity increase (see their Section 3.1.5), has little impact on the pre-MS radius evolution, we set $A_2=0$. We adopted the same input parameters (i.e., initial composition, mixing-length parameter, and overshooting parameter) as the K2 model with $A_2=0$ in \citet{Kunitomo2021}, which were optimized with solar observational constraints.

The evolution of the magnetospheric radius is described by the scaling of $r_{\rm mag}\propto R_*^{12/7}\dot{M}_{\rm acc}^{-2/7}$ under the assumption that $B_*$ and $M_*$ are constants with time. The stellar evolution model shows that
\begin{align}
    \frac{r_{\rm mag}}{R_*} \approx 4.3 F(t_{\rm age})
\end{align}
where $F(t_{\rm age})$ can be described as follows:
\begin{align}
    F(t_{\rm age}) &\approx \begin{dcases}
    1 & \text{for $0.3$~Myr $\lesssim t_{\rm age} < 1$~Myr}\\
    \left( \frac{t_{\rm age}}{1~{\rm Myr}}\right)^{0.190} & \text{for $t_{\rm age} \ge 1$~Myr}.
    \end{dcases}
\end{align}
The model shows that the value of $F(t_{\rm age})$ is approximately unity in $0.3~{\rm Myr}\lesssim t_{\rm age}\lesssim 1~{\rm Myr}$. We can derive a scaling relation for later evolution by using the relations of $R_\ast \propto t_{\rm age}^{-1/3}$ and $\dot{M}_{\rm acc}\propto t_{\rm age}^{-1.5}$.
Because of the weak dependence of $F(t_{\rm age})$ on the time, $r_{\rm mag}/R_\ast$ slowly changes with time and is of the order of unity in the time range of interest. Therefore, we approximate $r_{\rm mag}/R_*$ as a constant.

This study assumes that the stellar contraction occurs owing to the stellar radiation only. The conical disk wind also removes the energy from the protostar and enhances the stellar contraction \citep{Ferreira2000MNRAS}. However, considering the results of \citet{Ferreira2000MNRAS}, the stellar radius is insensitive to the effect (at most $\sim 10$\%, see their Figure~2). Therefore, our discussion based on the Kelvin-Helmholtz contraction will remain valid even with that effect.

\section{Scaling relations of spin-down time}\label{app:scaling_spin_down_time}

Using the definitions of $J_{\rm \ast,max}$ and $\dot{J}_{\rm CDW}$, we can show that $t_{\rm sd,up}$ scale as follows:
\begin{align}
    t_{\rm sd,up} &= \frac{0.5I_\ast \Omega_{\rm K}(R_\ast)}{f_A^2 \dot{M}_{\rm CDW}r_{\rm mag}^2\Omega_{\rm K}(r_{\rm mag})}\nonumber\\ 
    & \propto f_{\rm eff}^{-1} f_{A}^{-2}B_{\ast}^{-2/7}M_\ast^{15/14}R_\ast^{-5/14}\dot{M}_{\rm acc}^{-6/7}.
\end{align}
As the spin-down time $t_{\rm sd,up}$ only weakly depends on the stellar radius, deuterium burning before $\sim 0.5$~Myr has a minor impact on $t_{\rm sd,up}$.
At the pre-MS star stage, the stellar mass is nearly constant. If $f_{\rm eff}$, $f_{A}$, and $B_*$ do not significantly change during the evolution, the spin-down time mainly depends on $R_*$ and $\dot{M}$. In this case,
\begin{align}
    t_{\rm sd,up} \propto t_{\rm age}^{\frac{5}{42}+\frac{6}{7}a}=t_{\rm age}^{59/42}
\end{align}
for $a=3/2$. The gray dotted line in Figure~\ref{fig:spindown_time} indicates this scaling. This scaling is consistent with the results.

It is interesting to note that the spin-down time ($t_{\rm sd,up}$) decreases as the accretion rate increases if the accretion torque is always considerably smaller than the spin-down torque as considered in our model. In the actively accreting phase, the protostar shows a larger stellar radius and a higher mass loss rate of CDW. As the spin-down torque is an increasing function of the stellar radius and the wind mass loss rate, the protostar spins down efficiently in the early phase.

\section{Torques by the magnetospheric ejection and the stellar wind}\label{app:torques_ME_SW}
According to \citet{Gallet2019}, the torque by the magnetospheric ejection is written as
\begin{align}
    \dot{J}_{\rm ME} = K_{\rm ME}\frac{B_*^2 R_*^6}{r_{\rm mag}^3}\left[ K_{\rm rot}-\left( \frac{r_{\rm mag}}{r_{\rm cor}}\right)^{3/2}\right]
\end{align}
where $K_{\rm ME}=0.21$ and $K_{\rm rot}=0.7$ are the nondimensional parameters calibrated by the 2D MHD simulations of \citet{Zanni2013}. 
Note that within this formula, the torque represents a spin-down torque when its value is negative. As \citet{Gallet2019} indicated, the nondimensional calibration parameters are actually not fixed constants but vary as functions of the model parameters. Given the limited parameter range investigated by \citet{Zanni2013}, we align with their model by adopting a $r_{\rm cor}/r_{\rm mag}$ ratio similar to their results. In the propeller regime of their case C01, $r_{\rm cor}/r_{\rm mag}$ ranged approximately from 0.77 to 0.94 (see their Table~2). Based on this, we set $r_{\rm cor}/r_{\rm mag}=0.8$ as the standard value for Figure\ref{fig:torque}.

The spin-down torque by the stellar winds is expressed as
\begin{align}
    \dot{J}_{\rm SW} = \dot{M}_{\rm SW}r_{A}^2 \Omega_* \label{eq:torque_SW}
\end{align}
where $\dot{M}_{\rm SW}$ is the mass loss rate of the stellar wind and $r_A$ is the averaged Alfv\'en radius. Following \citet{Matt2012}, we write $r_{A}$ as
\begin{align}
    r_A = K_1 \left[ \frac{B_*^2 R_*^2}{\dot{M}_{\rm SW} \sqrt{K_2^2 v_{\rm esc}^2 + \Omega_*^2 R_*^2}}\right]^{m} R_* \label{eq:rA_Matt}
\end{align}
where $v_{\rm esc}=\sqrt{2GM_*/R_*}$ is the escape velocity, $m=0.2177$, $K_1=1.3$, and $K_2=0.0506$. Note that \citet{Gallet2019} adopt $K_1=1.7$, which is larger than the value given by \citet{Matt2012}. We also assume that the massive stellar wind is powered by accretion via unknown mechanisms and the efficiency is 10\%: $\dot{M}_{\rm SW}=f_{\rm SW}\dot{M}_{\rm acc}$ and $f_{\rm SW}=0.1$. Note that both the linear relation between $\dot{M}_{\rm SW}$ and $\dot{M}_{\rm acc}$ and the origin of the large efficiency are assumptions and remain elusive.


\bibliography{references}{}
\bibliographystyle{aasjournal}



\end{document}